\documentclass[prd,preprint,tightenlines,floatfix,showpacs,preprintnumbers,nofootinbib,eqsecnum,superscriptaddress]{revtex4}

 \usepackage[dvips,final]{graphicx}
  \usepackage{amssymb}
    \usepackage{amsfonts}
     \usepackage{epsfig}
	\usepackage{subfigure}
      \usepackage{bm}

\def\Pom{{\bf I\!P}}
\def\Reg{{\bf I\!R}}

\newcommand{\bkappa}{\mbox{\boldmath $\kappa$}}
\newcommand{\bp}{\mbox{\boldmath $p$}}
\newcommand{\bq}{\mbox{\boldmath $q$}}

\newcommand{\bk}{\mbox{\boldmath $k$}}
\newcommand{\bb}{\mbox{\boldmath $b$}}

\newcommand{\bM}{\mbox{\boldmath $M$}}

\newcommand{\ket}[1]{| {#1} \rangle}
\newcommand{\bra}[1]{\langle {#1} |}

\newcommand{\half}{{1\over 2}}
\def\lsim{\mathrel{\rlap{\lower4pt\hbox{\hskip1pt$\sim$}}
    \raise1pt\hbox{$<$}}}         
\def\gsim{\mathrel{\rlap{\lower4pt\hbox{\hskip1pt$\sim$}}
    \raise1pt\hbox{$>$}}}         

\begin{document}

\thispagestyle{empty} \preprint{\hbox{}} \vspace*{-10mm}

\title{Production of $Z^0$ bosons with rapidity gaps: \\
exclusive photoproduction 
in $\gamma p$ and $p p$ collisions \\
and inclusive double diffractive $Z^0$'s}

\author{A. Cisek}
\email{Anna.Cisek@ifj.edu.pl}
\affiliation{Institute of Nuclear Physics PAN, PL-31-342 Cracow,
Poland} 
\author{W. Sch\"afer}
\email{Wolfgang.Schafer@ifj.edu.pl}
\affiliation{Institute of Nuclear Physics PAN, PL-31-342 Cracow,
Poland} 
\author{A. Szczurek}
\email{Antoni.Szczurek@ifj.edu.pl}
\affiliation{Institute of Nuclear Physics PAN, PL-31-342 Cracow,
Poland} 
\affiliation{University of Rzesz\'ow, PL-35-959 Rzesz\'ow,
Poland}

\date{\today}

\begin{abstract}
We extend the $k_\perp$--factorization
formalism for exclusive 
photoproduction of vector mesons to the production
of electroweak $Z^0$ bosons.
Predictions for the $\gamma p \to Z^0 p$ and
$p p \to p p Z^0$ reactions are given using
an unintegrated gluon distribution tested against
deep inelastic data. 
We present distributions in the $Z^0$ rapidity, transverse
momentum of $Z^0$ as well as in relative azimuthal angle
between outgoing protons.
The contributions of different flavours are discussed. 
Absorption effects lower the cross section
by a factor of 1.5-2, depending on the Z-boson rapidity.
We also discuss the production of $Z^0$ bosons in central
inclusive production.
Here rapidity and $(x_{\Pom,1}, x_{\Pom,2})$ distributions
of $Z^0$ are calculated.
The corresponding cross section is about three orders of magnitude 
larger than that for the purely exclusive process.
\end{abstract}


\maketitle

\section{Introduction}

There has been recently much experimental progress
in the field of central exclusive production. 
The observation of exclusive central dijets \cite{exclusive_dijets},
as well as charmonia/$\mu^+ \mu^-$--pairs\cite{exclusive_charmonia} 
has clearly demonstrated the feasibilty of detecting exclusive
final states at collider energies.
These are good prospects for possible future studies 
at the LHC, addressing a wide range of physics problems, 
from Higgs physics, to the investigations of the QCD--Pomeron and
hadronic structure of the produced particles. For recent
reviews see for example \cite{Martin_epiphany} for theory/phenomenology,
and \cite{experiment} for experiment.

\begin{figure}[!htb] %
\begin{center}
\includegraphics[height=4.5cm]{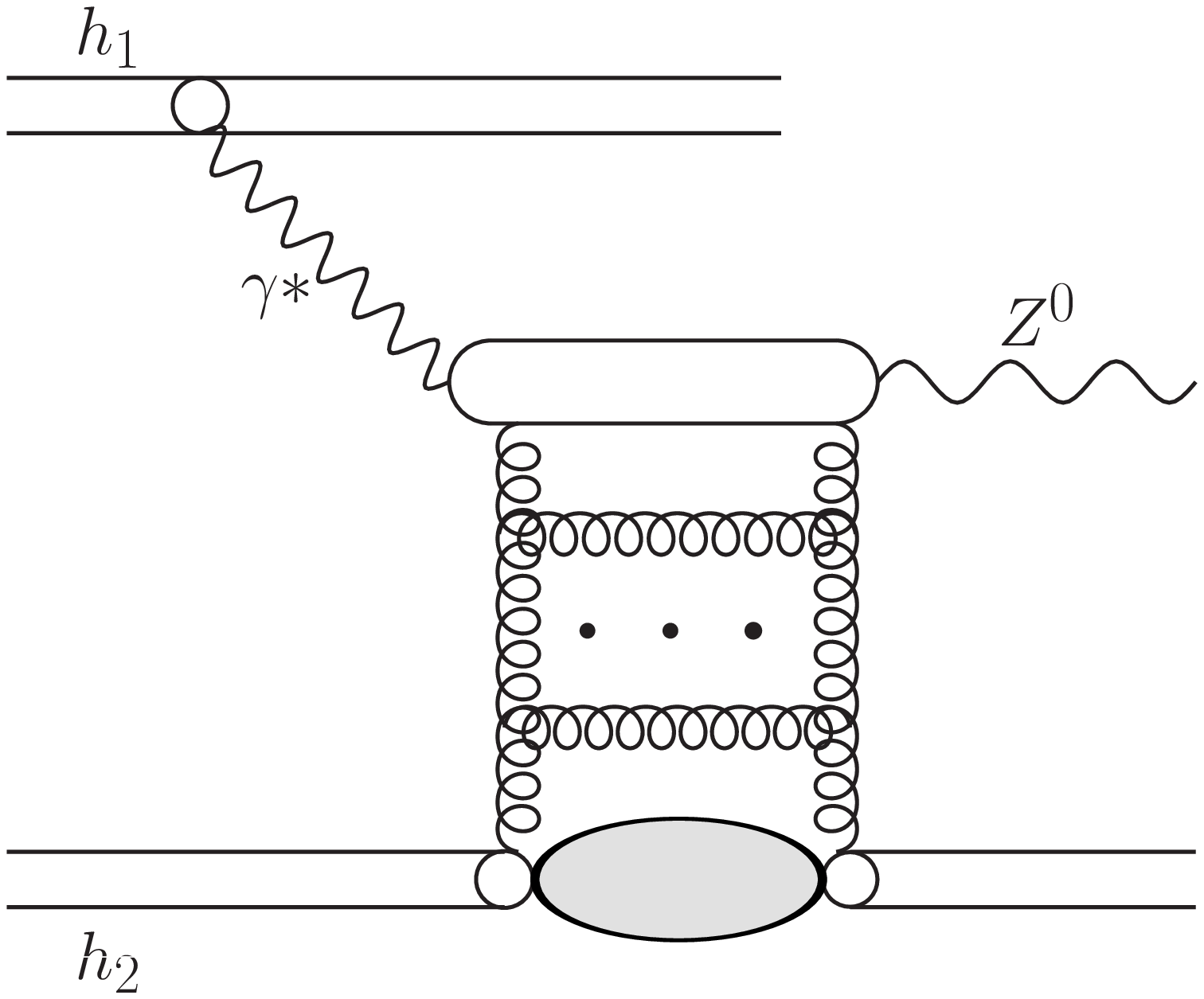}
\includegraphics[height=4.5cm]{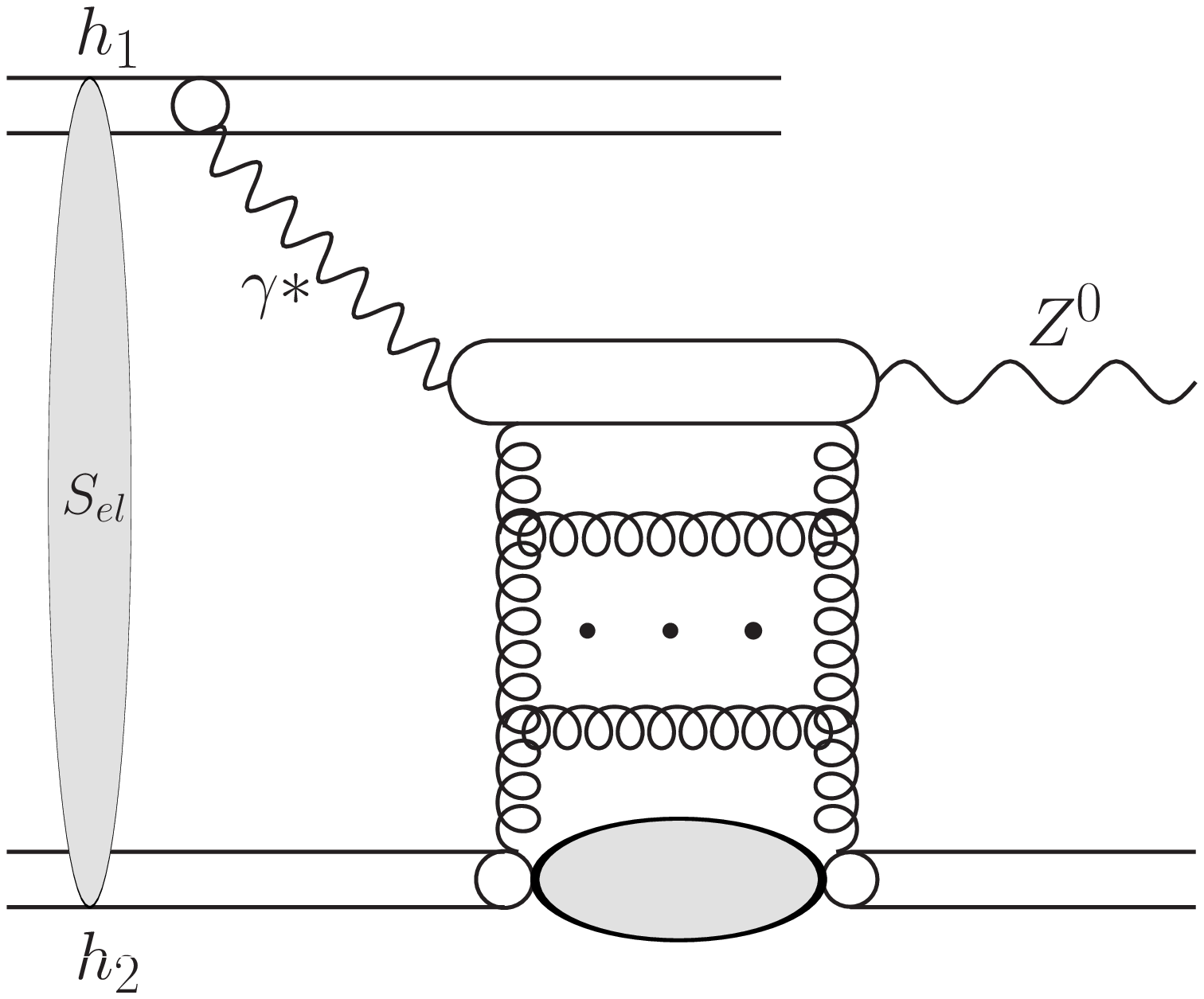}
\caption[*]{Diagrammatic representation of the photon--Pomeron
fusion mechanism of exclusive $Z^0$--boson production 
in hadronic collisions.
\label{fig:Born_diagrams}
}
\end{center}
\end{figure}

The mechanism of the reaction strongly depends on the
centrally produced particle, in particular its spin, 
parity, C-parity and internal stucture. 
For heavy vector quarkonia such as
$J/\Psi$ and $\Upsilon$ the photon-pomeron fusion is
the dominant mechanism (for recent calculations
see e.g. \cite{SS07,RSS08}). 
The same is expected for the $Z^0$ gauge boson 
\cite{GM08,MW08}. The dominant reaction mechanism
is shown in Fig.\ref{fig:Born_diagrams}.


Here, the essential ingredient is the $\gamma p \to Z^0 p$
subprocess, which proceeds through the $q\bar q$ component
of the virtual photon. There is a strong similarity to the 
production of $q \bar q$ vector mesons, and one would expect
the QCD description of this process to follow from
the color dipole/$k_\perp$-factorization approach 
to vector meson production (for a review, see \cite{INS06}) 
by straightforward modifications \cite{GM08,MW08}.
The important distinction to the case of vector meson 
production is the fact that the $q \bar q$ pair 
coupling to the $Z^0$--boson can be put on its mass-shell.  
In the impact parameter space color-dipole formulation, this
requires to continue the light--cone wavefunction of the 
$Z^0$ to the region of complex arguments. The 
resulting highly oscillatory integrands are however
not straightforward to handle \cite{MW08}.

In the momentum space representation given in this work,
the situation is more transparent, and the numerics poses
no special problems. 

Previous calculations of the $pp \to pp Z^0$ process
made use of the equivalent-photon approximation (EPA) 
and did not include absorption effects.
In the EPA only total cross section or rapidity 
distribution of the $Z^0$ boson can be calculated.
In this paper we use the formalism of \cite{SS07,RSS08} 
to perform the calculation at the amplitude--level,
which allows us to
calculate other differential observables (e.g. in $Z^0$
boson transverse momentum or correlation in relative 
azimuthal angle between outgoing protons) and 
to include absorption corrections.

The cross sections for exclusive $Z^0$ production
for both the Tevatron and LHC are very small. 
In fact a recent search for exclusive $Z^0$\cite{CDF_Z0_exclusive}
only puts rather generous bounds on the cross section.
The exclusive events are characterized by large rapidity gaps between 
centrally produced $Z^0$ bosons and very
forward or very backward final state nucleons.
Another process with these features is the inclusive double-diffractive 
production of $Z^0$, which to our knowledge was not 
previously calculated the literature 
\footnote{So far only the single-diffractive
contribution was estimated in the literature,
see e.g. \cite{Single_Diffraction}.}
.
In the latter case the $Z^0$ in 
the central rapidity region is associated with 
low-multiplicity hadronic activity.

\section{$\gamma p \to Z^0 p$ photoproduction process}


\begin{figure}[!htb] %
\begin{center}
\includegraphics[height=5.5cm]{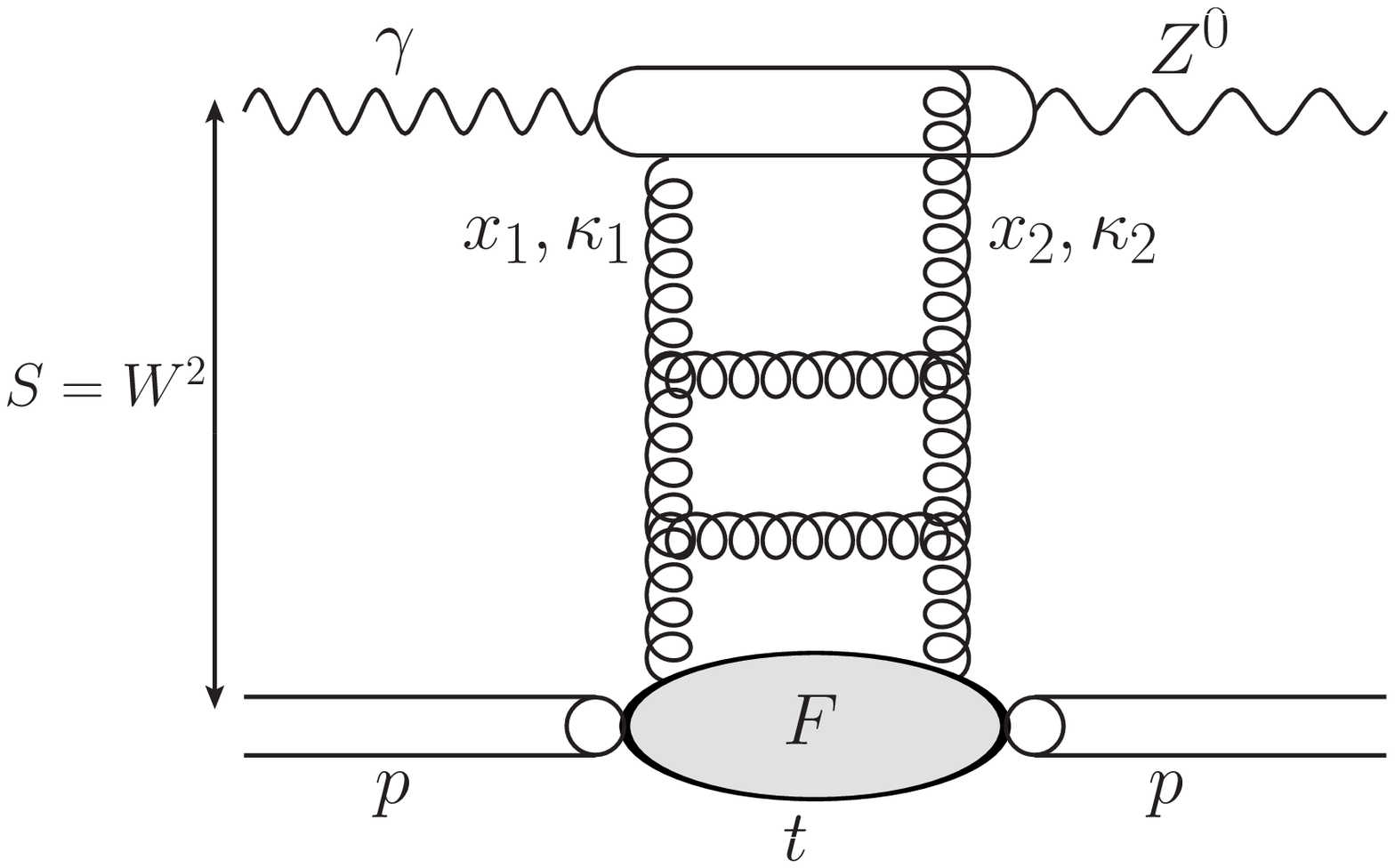}
\caption[*]{A pQCD diagram for the $\gamma p \to Z^0 p$ amplitude
at large $\gamma p$ cm-energy. The gluon exchange ladder is modelled
by the unintegrated gluon distribution of the target.
\label{fig:photoZ_diagram}
}
\end{center}
\end{figure}


Before we go to the hadronic reaction let us start
from the real photoproduction process depicted in
Fig.\ref{fig:photoZ_diagram}.
The forward production amplitude
can be written very much the same way as for exclusive
photoproduction of vector quarkonia (see for example \cite{INS06}):
\begin{eqnarray}
{\cal M}(W,\Delta^2 = 0) =
W^2 \, \ \sum_{f} {4 \pi \alpha g_V^f \over 4\pi^2}
\, 2 \int_0^1 dz
\int d^2\bk { {\cal{A}}_f (z,\bk^2) \over  \bk^2 + m_{f}^{2} - z(1-z) M_Z^2 - i \epsilon} \; .
\label{forward_photoproduction_amplitude}
\end{eqnarray}
As was already shown in \cite{GM08,MW08}, only the vectorial 
part of the $Z^0 q \bar q$--coupling contributes to 
the forward amplitude, 
\begin{equation}
g_V^f = { I_{3f} - 2 e_{f} \sin^{2}\Theta_{W} 
\over  \sin 2 \Theta_{W}}
\end{equation}
is the relevant weak vector coupling, $I_{3f}$ is the weak 
isospin of a quark of flavour $f$, $e_f$ is its charge,
and $\Theta_W$ is the Weinberg angle.
The imaginary part of ${\cal{A}}_f$ can be obtained from the 
results given for vector meson production with the 
$\gamma_\mu$ vertex in \cite{INS06}. Performing azimuthal 
integrals one obtains \cite{RSS08}:
\begin{equation}
\Im m {\cal A}_f(z,\bk^2) = \pi 
\int_0^\infty
{\pi d\kappa^2 \over \kappa^4} \alpha_S(q^2) {\cal{F}}(x,\kappa^2)
(  A_{0f}(z,k^2) \; W_{0f}(k^2,\kappa^2) + A_{1f}(z,k^2) \; W_{1f}(k^2,\kappa^2) )
\end{equation}
with
\begin{eqnarray}
A_{0f}(z,k^2) = m_f^2 \, \, , \, \, 
A_{1f}(z,k^2) = [ z^2 + (1-z)^2] \frac{k^2}{k^2+m_f^2} \, ,
\nonumber
\end{eqnarray}
and 
\begin{eqnarray}
W_{0f}(k^2,\kappa^2) &=& 
{1 \over k^2 + m_f^2} - {1 \over \sqrt{(k^2-m_f^2-\kappa^2)^2 + 4 m_f^2 k^2}} \; ,
\nonumber \\
W_{1f}(k^2,\kappa^2) &=& 1 - { k^2 + m_f^2 \over 2 k^2}
\left( 1 + {k^2 - m_f^2 - \kappa^2 \over 
\sqrt{(k^2 - m_f^2 - \kappa^2)^2 + 4 m_f^2 k^2 }}
 \, \right) \; .
\nonumber
\end{eqnarray}
The strong coupling $\alpha_s$ enters at the scale 
$q^2 = \max\{\bk^2 + m_f^2,\bkappa^2\}$. The unintegrated gluon distribution
${\cal{F}}(x,\bkappa^2)$ is sampled at  
$x = c_{skewed}\cdot M_Z^2/W^2$ with $c_{skewed} = 0.41$.
This shifted value of $x$ simulates the prescription of
\cite{Shuvaev} to obtain the skewed distribution from the 
diagonal one, and is valid for the particular gluon distribution
we use \cite{INS06}.
Now, let
\begin{equation}
z_\pm = {1 \over 2} \left( 1 \pm \sqrt{1 - {4m_{f}^2 \over M_Z^2}} \right) \, ,
\end{equation}
then, the integration domain $z \in [0,1]$ must be split into
$[0,z_-] \cup [z_-,z_+] \cup [z_+,1]$. 
Apparently, within the subdomain $[z_-,z_+]$ the denominator in
Eq.(\ref{forward_photoproduction_amplitude}) can go to zero, 
which means that the $q\bar q$ state after the interaction 
can go on-shell. 
This leads to a rotation of the complex phase of 
the dominantly imaginary amplitude.
For $z \in [z_-,z_+]$, one must use the Plemelj-Sokhocki formula
\begin{eqnarray}
{1 \over \bk^2 + m_{f}^{2} - z(1-z) M_Z^2 - i \epsilon}
= \mathrm{PV}
{1 \over \bk^2 + m_{f}^{2} - z(1-z) M_Z^2 } + \, i \, \pi \, 
\delta(   \bk^2 + m_{f}^{2} - z(1-z) M_Z^2  )
\, , \nonumber \\
\end{eqnarray}
where PV denotes the principal value integral. It can be evaluated as
\begin{eqnarray}
\mathrm{PV} \, \int_0^\infty 
d\bk^2 {{\cal A}_{f}(z,\bk^2) \over \bk^2 - \tau^2}
&=&
\int_0^{(1+\lambda)\tau^2}
d\bk^2 {{\cal A}_{f}(z,\bk^2) - 
{\cal A}_{f}(z,\tau^2) \over \bk^2 - \tau^2}
+ \int_{(1+\lambda)\tau^2}^\infty 
d\bk^2 {{\cal A}_{f}(z,\bk^2) \over \bk^2 - \tau^2}
\nonumber \\
&+& {\cal A}_{f}(z,\tau^2) \, \log(\lambda) \, ,
\nonumber \\
\end{eqnarray}
for an arbitrary positive value of $\lambda$.
Here $\tau^2 = z(1-z)M_Z^2 - m_{f}^2$ is positive in the relevant 
integration domain.

Another distinction in comparison to the vector-meson(VM) photoproduction
is worth a comment. Effectively, we replace the non--perturbative
light cone wave--function of the VM by the propagator:
\begin{eqnarray}
\psi_V(z,\bk^2) \to {1 \over \bk^2 + m_f^2 - z(1-z)M_Z^2 - i \epsilon}
\, .
\end{eqnarray}
While in the case of vector--mesons, the light--cone wave--function
will suppress the endpoint--region $z,1-z \ll 1$, no such suppression
of asymmetric $q \bar q$ pairs is available here.
Incidentally, in impact parameter space asymmetric pairs correspond
to large dipole sizes \cite{NZ91},  and it is precisely the
wave--function suppresion of large dipoles, which leads
to the dipole--size scanning property \cite{KNNZ93,INS06} of VM production 
amplitudes.  
Therefore, strictly speaking, the $Z^0$ production cross section
is not purely perturbatively calculable, but one must rely on the
ability of the color--dipole/$k_\perp$-factorization approaches
to properly factorize the large dipole/infrared contributions.
Compare this to the {\emph{scaling}} contribution of large
dipoles to the transverse DIS structure function 
$F_T(x,Q^2) = 2x F_1(x,Q^2)$ \cite{NZ91,IN_glue} at large $Q^2$.

Finally, we note, that we restore the real part of the amplitude
by substituting
\begin{eqnarray}
{\cal{A}}_f = (i + \rho) \Im m {\cal{A}}_f \, ,
\end{eqnarray}
where $\rho = \tan (\pi \Delta_\Pom/2)$, and 
$\Delta_\Pom = \partial \log \Im m {\cal{A}}_f /\partial \log W^2$,
and the $\gamma p \to Z^0 p$ amplitude within the diffraction 
cone is given by
\begin{eqnarray}
{\cal{M}}(W^2,\Delta^2) = {\cal{M}}(W^2,\Delta^2=0) \, \exp[Bt] \, ,
\label{eq:t-dependence}
\end{eqnarray}
where $t = -\Delta^2$ and the running diffraction slope 
is taken as
\begin{eqnarray}
B = B_0 + 2 \alpha_{eff}' \log\Big(W^2 / W_0^2 \Big) \, ,
\label{slope_exclusive}
\end{eqnarray}
with $B_0 = 3.5 \, \mathrm{GeV}^{-2}$, 
$\alpha'_{eff} = 0.164 \, \mathrm{GeV}^{-2}$, and 
$W_0 = 95 \, \mathrm{GeV}$ \cite{H1_JPsi}.

\section{$p p \to p p Z^0$ exclusive hadroproduction}

Assuming only helicity conserving processes
the Born amplitude for the $p p \to p p Z^0$ reaction
is a sum of amplitudes of the processes shown in
Fig.\ref{fig:Born_diagrams} and can be written through 
the amplitudes of photoproduction processes 
$\gamma h_2 \to Z^0 h_2$ or $\gamma h_2 \to Z^0 h_2$, 
discussed above, in the form of the vector
\begin{eqnarray}
{\bM}^{(0)}_{h_1 h_2 \to h_1 h_2 Z^0}(\bp_1,\bp_2)
&=& e_1 \sqrt{4 \pi \alpha_{em}} F_1(t_1) \frac{2 {\bp}_1}{z_1 t_1}
\cdot {\cal M}_{\gamma^* h_2 \to Z^0 h_2}(Q_1^2;s_2,t_2)
\nonumber \\
&+& e_2 \sqrt{4 \pi \alpha_{em}} F_1(t_2) \frac{2 {\bp}_2}{z_2 t_2}
\cdot {\cal M}_{\gamma^* h_1 \to Z^0 h_1}(Q_2^2;s_1,t_1) \; .
\label{pp_ppZ_amplitude}
\end{eqnarray}
Above $Q_1^2 = -t_1$ and $Q_2^2 = -t_2$ are virtualities
of photons, $F_1$ is the familiar Dirac electromagnetic 
form factor of the proton/antiproton and $\bp_1$,
$\bp_2$ are transverse momenta of outgoing protons.
In the present analysis we shall use a simple parametrization 
of the Dirac electromagnetic form factor $F_1$ taken from
Ref.\cite{DDLN02}.

The dependence of the the virtual photoproduction subprocess
amplitude on space-like virtuality of the photon $Q^2$
is in practice entirely negligible (recall that $Q^2 \ll M_Z^2$).

\begin{figure}[!htb] %
\begin{center}
\includegraphics[height=4.5cm]{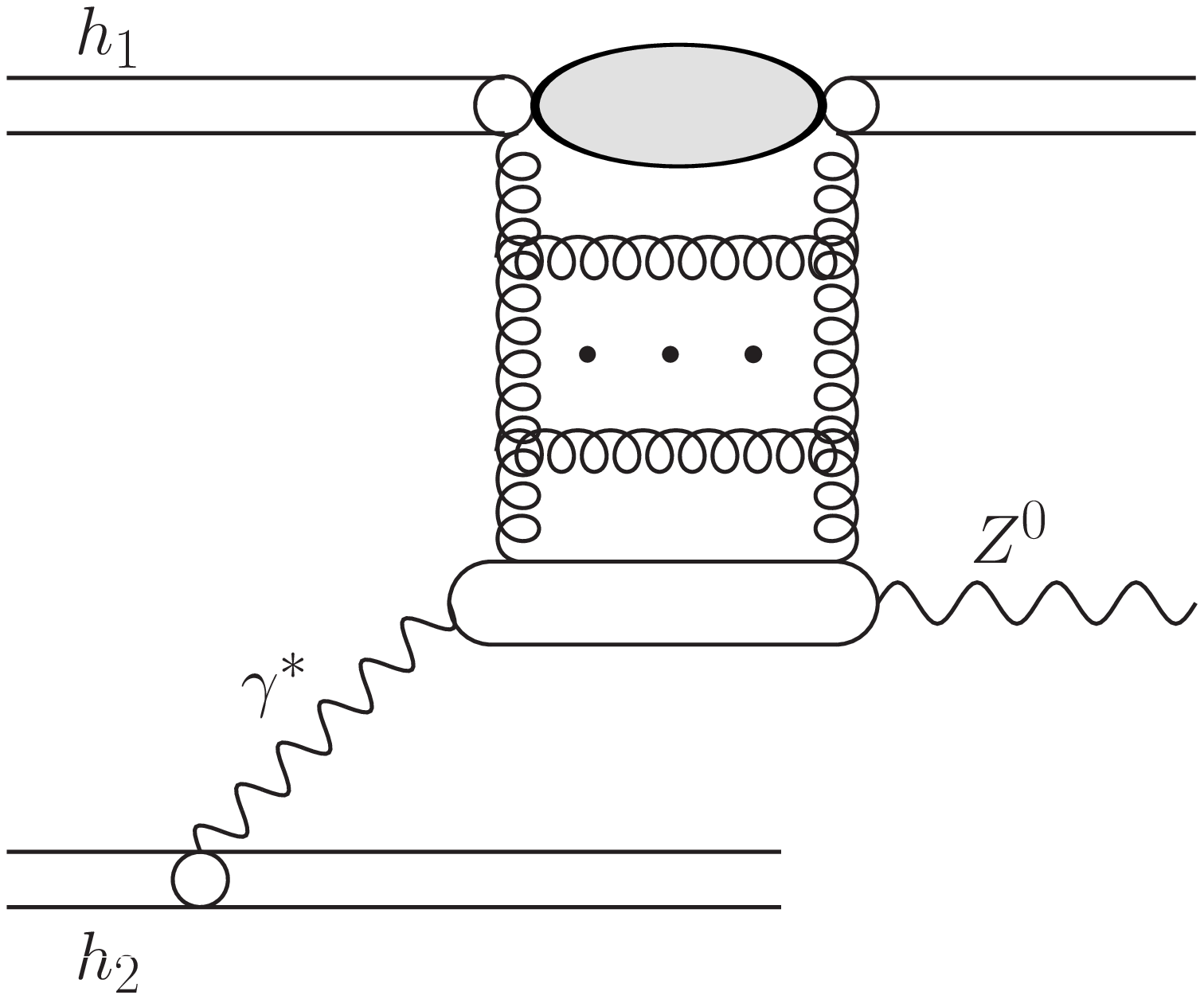}
\includegraphics[height=4.5cm]{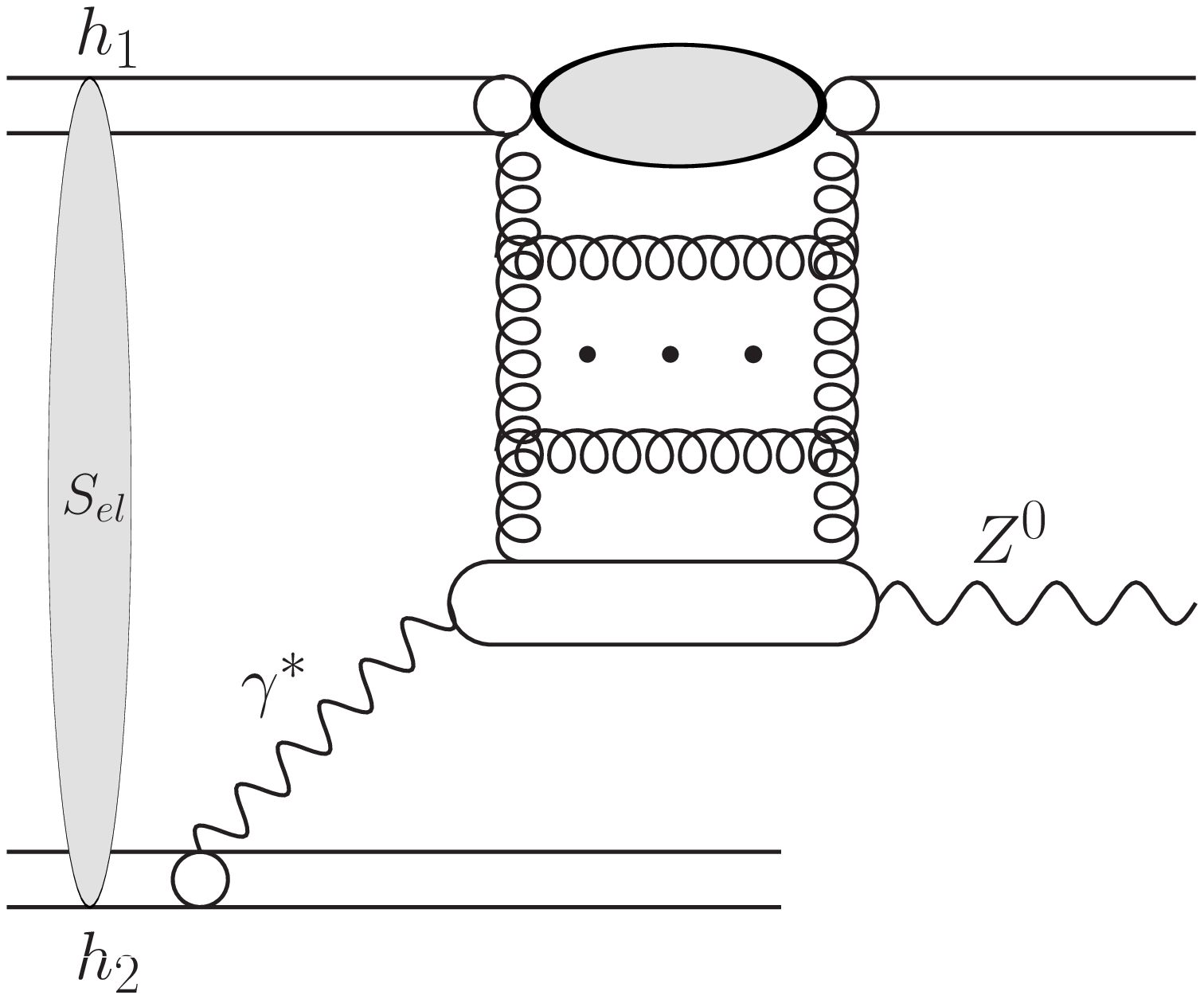}
\caption[*]{Diagrammatic representation of the 
exclusive $Z^0$ production amplitude with inclusion 
of absorptive corrections. The absorptive corrections
are calculable in terms of the hadron--hadron elastic $S$-matrix.
\label{fig:diagrams_absorption}
}
\end{center}
\end{figure}


In the hadroproduction process
one has to include additional absorption corrections.
The relevant formalism for the calculation
of amplitudes and cross--sections was reviewed in 
some detail in Ref.\cite{SS07}. Here we give only 
a the main formulas. 
The basic mechanisms are shown in 
Fig.\ref{fig:diagrams_absorption}. 

Inclusion of absorptive corrections (the 'elastic rescattering')
leads in momentum space to the full, absorbed amplitude
\begin{eqnarray}
\bM(\bp_1,\bp_2) &&= \int{d^2 \bk \over (2 \pi)^2} \, S_{el}(\bk) \,
\bM^{(0)}(\bp_1 - \bk, \bp_2 + \bk)  
= \bM^{(0)}(\bp_1,\bp_2) - \delta \bM(\bp_1,\bp_2) \, .
\nonumber \\
\label{rescattering term}
\end{eqnarray}
With 
\begin{equation}
S_{el}(\bk) = (2 \pi)^2 \delta^{(2)}(\bk) - \half T(\bk) \, \, \, ,
\, \, 
\, T(\bk) = 
\sigma^{p \bar p}_{tot}(s) \, \exp\Big(-\half B_{el} \bk^2 \Big) \, ,
\end{equation}
where at Tevatron energy $\sqrt{s}$ = 1800 GeV 
$\sigma^{p \bar p}_{tot}(s) = 76$ mb, $B_{el} = 17 $ GeV$^{-2}$
\cite{CDF_total} ,
the absorptive correction $\delta \bM$ reads 
\begin{eqnarray}
\delta \bM(\bp_1,\bp_2) = \int {d^2\bk \over 2 (2\pi)^2} \, T(\bk) \,
\bM^{(0)}(\bp_1-\bk,\bp_2+\bk) \, .
\label{absorptive_corr}
\end{eqnarray}

The differential cross section is given in terms of $\bM$ as
\begin{equation}
d \sigma = { 1 \over 512 \pi^4 s^2 } | \bM |^2 \, dy dt_1 dt_2
d\phi \, ,
\label{differential_cross_section}
\end{equation}
where $y$ is the rapidity of the 
$Z^0$-boson, $t_1$, $t_2$ are four-momentum 
transfers squared, and $\phi$ is the angle between transverse
momenta $\bp_1$ and $\bp_2$.

\section{Inclusive double diffactive production of $Z^0$}

\begin{figure}[!htb] %
\begin{center}
\includegraphics[height=5.5cm]{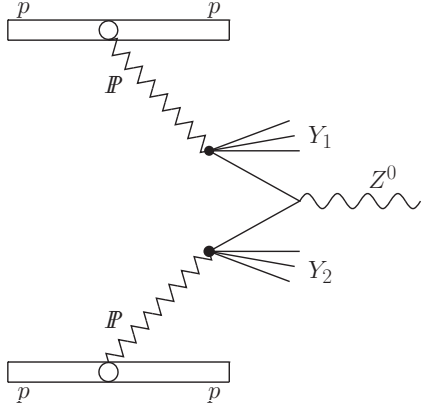}
\caption[*]{Pomeron--Pomeron fusion representation of the 
central diffractive inclusive $Z^0$ production. 
The hard subprocess is viewed as a collision of partons of the
Pomeron.
\label{fig:double_pomeron_diagram}
}
\end{center}
\end{figure}

The purely exclusive process discussed so far 
can be characterized by large rapidity gaps
between the centrally produced $Z^0$
and forward/backward emitted protons/antiprotons.
Can the rapidity gap method be used to identify this
process? The discussed exclusive process is not 
the only one with double rapidity gaps.
The inclusive double-pomeron cross section 
for $Z^0$ boson production was apparently not 
previously calculated.
The mechanism is depicted in 
Fig.\ref{fig:double_pomeron_diagram}.


Following Ingelman and Schlein \cite{Ingelman_Schlein}, 
one may try to estimate the
hard diffractive process by assuming that the Pomeron has a
well defined partonic structure, and that the hard process
takes place in a Pomeron--Pomeron collision.
Then the rapidity distribution of $Z^0$--bosons would
be obtained from
\begin{eqnarray}
{d \sigma(pp \to pp Z^0 X) \over dy} = K \, \sum_f
\sigma(q_f \bar q_f \to Z^0; x_1 x_2 s) \, \Big( x_1 q_f^D(x_1,\mu^2)
\, x_2 \bar q_f^D(x_2,\mu^2) +  (1\leftrightarrow 2) \Big) \, .
\nonumber \\ 
\label{naive_factorization}
\end{eqnarray}
Here
\begin{eqnarray}
x_1 = {M_Z \over \sqrt{s}} e^y , \, \, \, x_2 = {M_Z \over \sqrt{s}} e^{-y} , 
\nonumber
\end{eqnarray}
$\sigma_{q_f {\bar q}_f \to Z^0}({\hat s})$ is the
elementary ``zeroth-order'' flavour-dependent cross sections
(see e.g. \cite{BP87}). $K$ in Eq.(\ref{naive_factorization}) 
stands for the Drell-Yan type $K$-factor which includes 
approximately pQCD NLO corrections \cite{BP87}. 

The effective 'diffractive' quark distribution of
flavour $f$ is given by a convolution of the flux of Pomerons
$f_\Pom(x_\Pom)$ and the parton distribution in a Pomeron 
$q_{f/\Pom}(\beta, \mu^2)$:
\begin{eqnarray}
q_f^D(x_1,\mu^2) = \int d x_\Pom d\beta \, \delta(x-x_\Pom \beta) 
q_{f/\Pom} (\beta,\mu^2) \, f_\Pom(x_\Pom) \, 
= \int_x^1 {d x_\Pom \over x_\Pom} \, f_\Pom(x_\Pom)  
q_{f/\Pom}({x \over x_\Pom}, \mu^2) \, . \nonumber \\
\end{eqnarray}
The flux of Pomerons $f_\Pom(x_\Pom)$ enters in the form integrated over 
four--momentum transfer 
\begin{eqnarray}
f_\Pom(x_\Pom) = \int_{t_{min}}^{t_{max}} dt \, f(x_\Pom,t) \, ,
\end{eqnarray}
with $t_{min}, t_{max}$ being the kinematic boundaries.

Both pomeron flux factors $f_{\Pom}(x_{\Pom},t)$ as well 
as quark/antiquark distributions in pomeron were taken from 
the H1 collaboration analysis of diffractive structure function
and/or from the analysis of diffractive dijets at HERA
\cite{H1}. The factorization scale is taken as
$\mu_F^2 = M_Z^2$.

\begin{figure}[!htb] 
\begin{center}
\subfigure[]{\label{a}
\includegraphics[height=6cm]{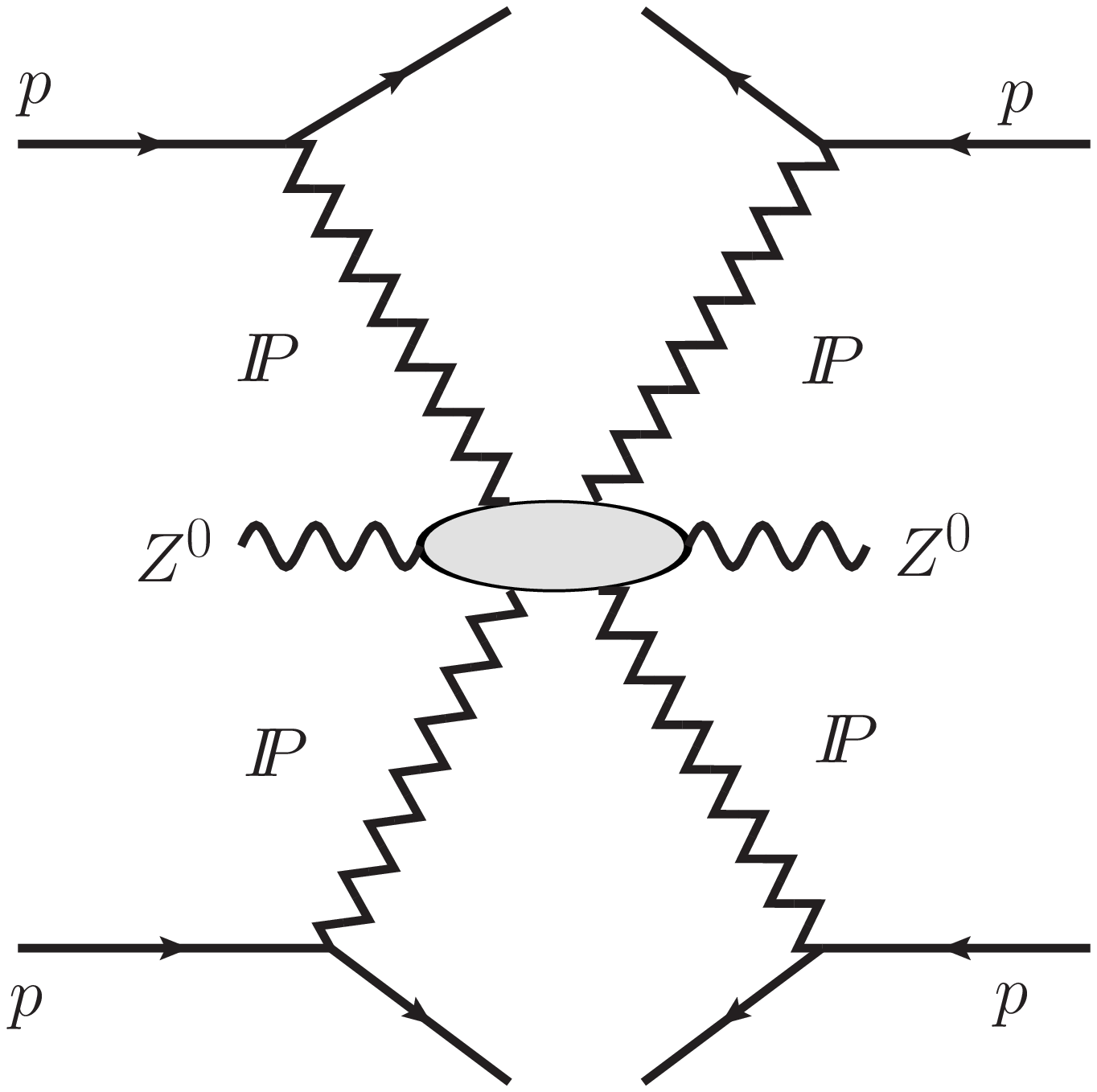}}
\subfigure[]{\label{b}
\includegraphics[height=6cm]{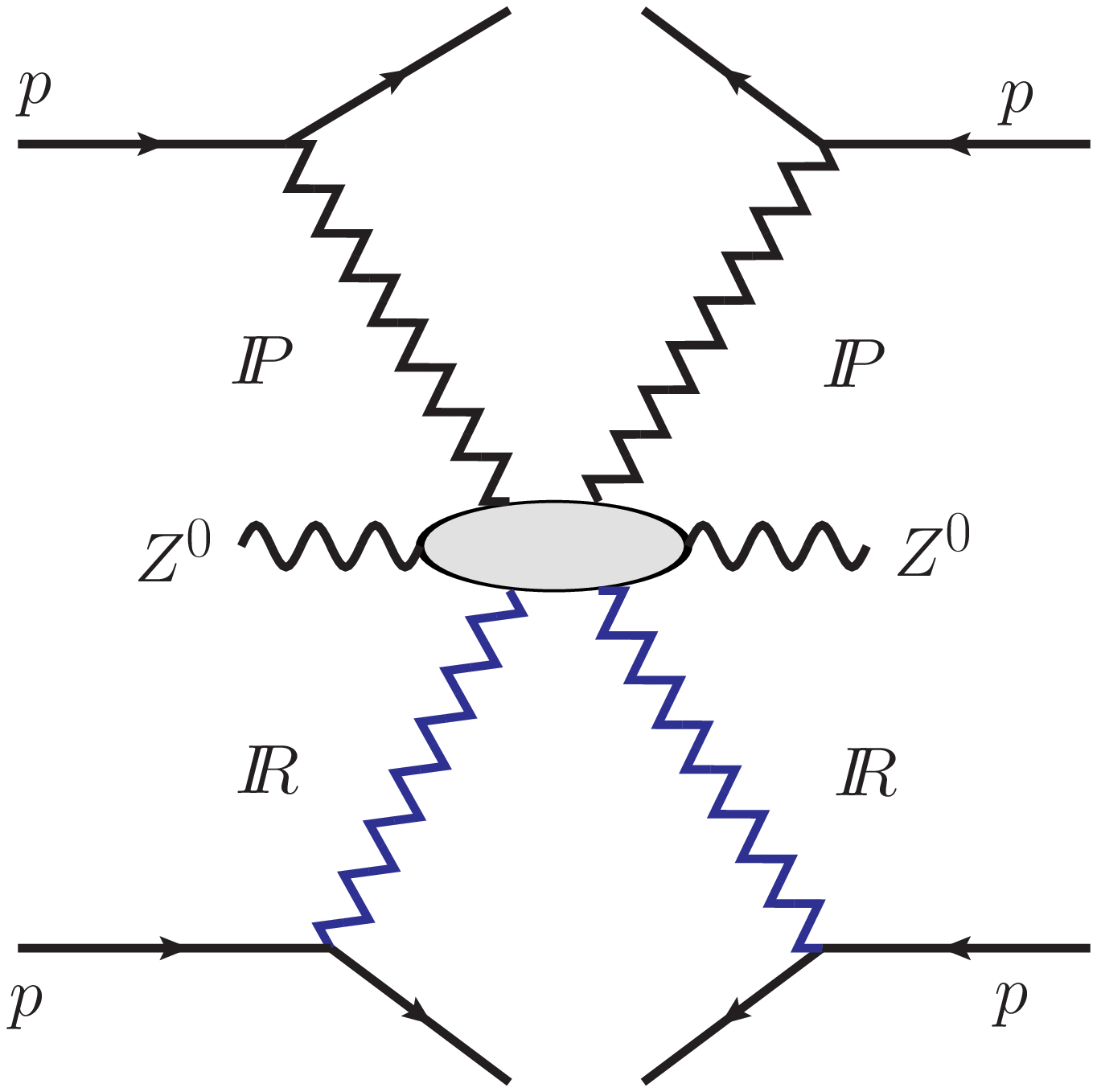}}
\subfigure[]{\label{c}
\includegraphics[height=6cm]{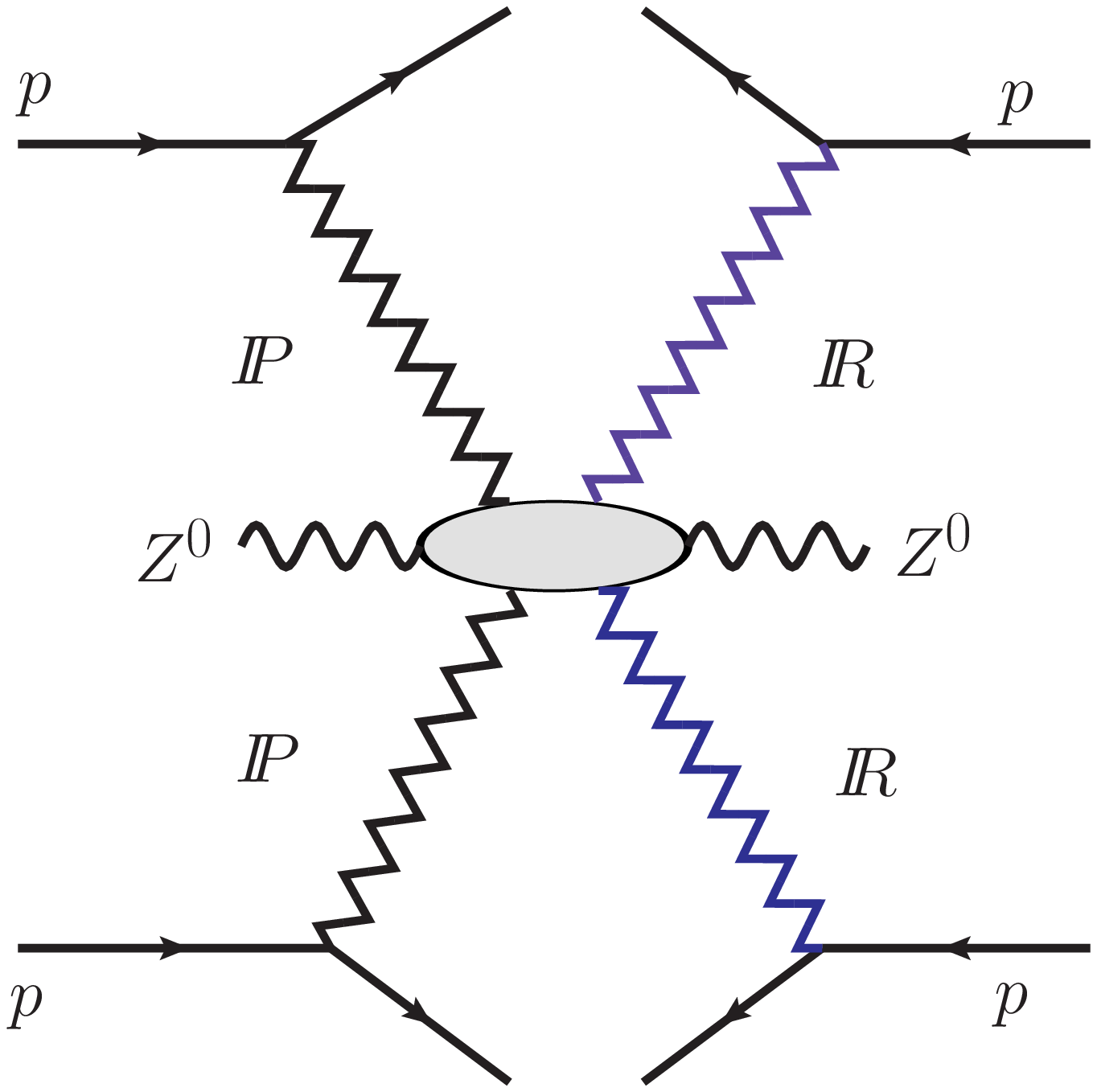}}
\subfigure[]{\label{d}
\includegraphics[height=6cm]{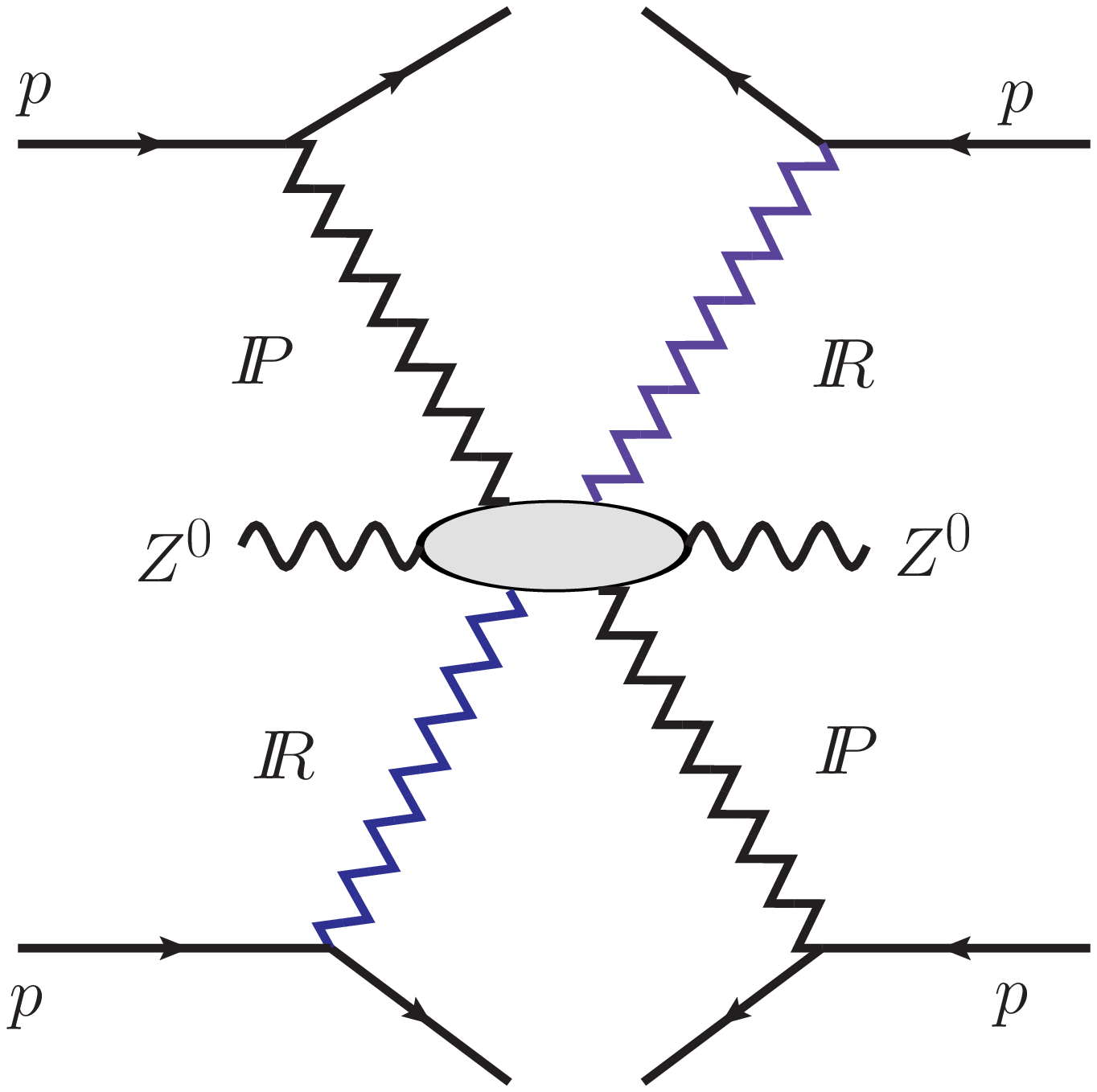}}
\subfigure[]{\label{d}
\includegraphics[height=6cm]{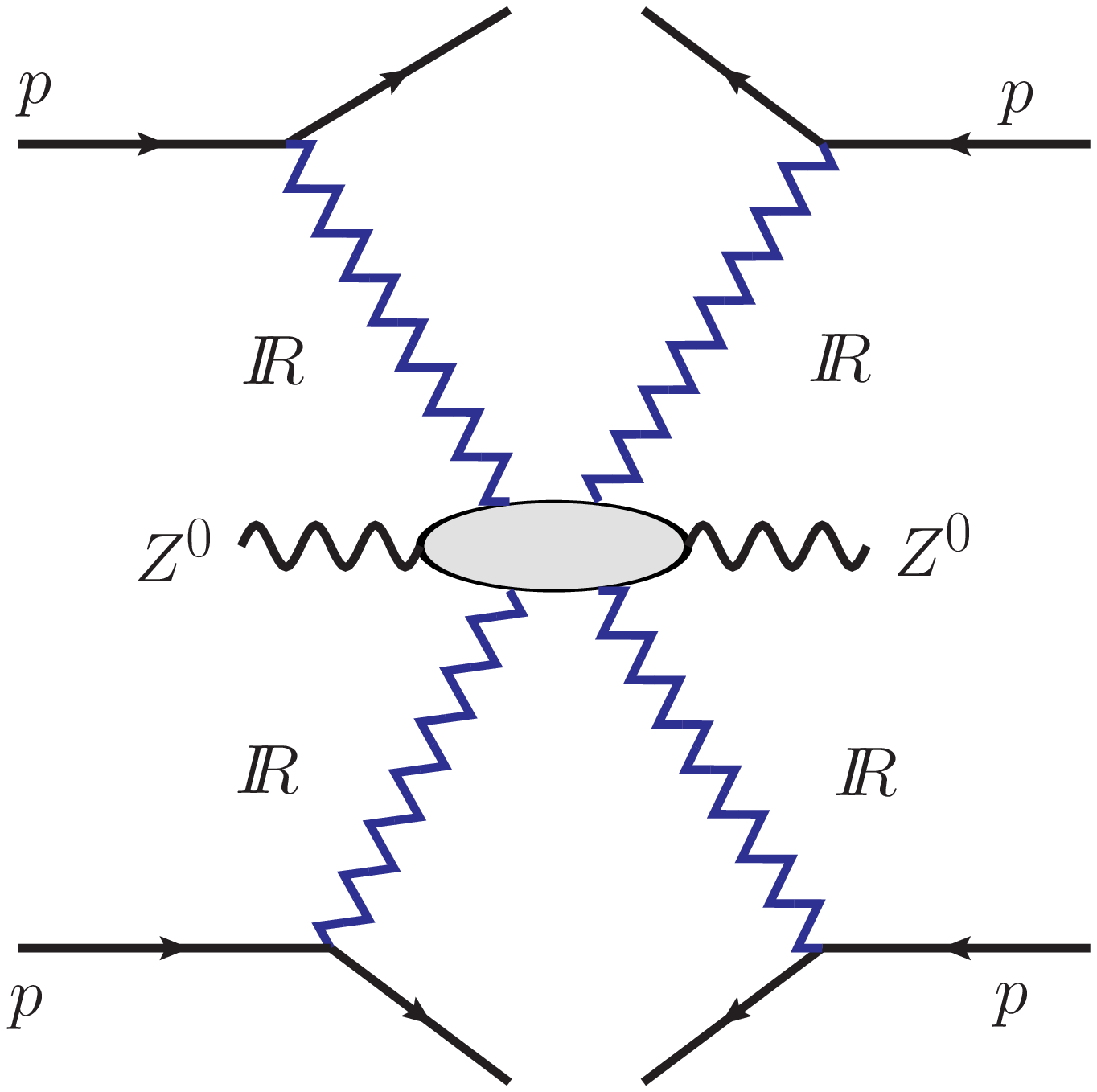}}
\caption[*]{Mueller-Kancheli diagrams for the process
$pp \to p + \mathrm{gap} + Z^0 X + \mathrm{gap} + p$.
Interference diagrams of type (c,d), which would involve
$\Pom-\Reg$ interference structure functions are neglected.
\label{fig:diagrams}
}
\end{center}
\end{figure}

Besides the Pomeron--exchange, one must also include the
secondary Reggeon--exchange contribution, which dominates
at larger $x_\Pom$. Consequently, a number of interference
contributions arise, which are shown diagramatically in Fig.
\ref{fig:diagrams}. As we wish to use the results of the 
H1 analysis of diffractive DIS at HERA, we have to omit a number
of interference terms, which would involve Reggeon--Pomeron interference
structure functions that have been neglected in the H1 analysis.
Incidentally, within pQCD, there is no reason why interference 
terms should be small, and in fact, they enter the 
diffractive structure functions in the maximal possible way \cite{WS98}.
While one obtains good fits of HERA data, even 
omitting the $\Reg-\Pom$ interference, the rather unphysical values 
of the so-extracted Reggeon trajectory parameters show the limitations
of such a procedure. Clearly though, a full reanalysis of H1-data is not
warranted for our purpose of obtaining a first estimate of the
$Z^0$-production cross section.


\begin{figure}[!htb] %
\begin{center}
\includegraphics[height=6cm]{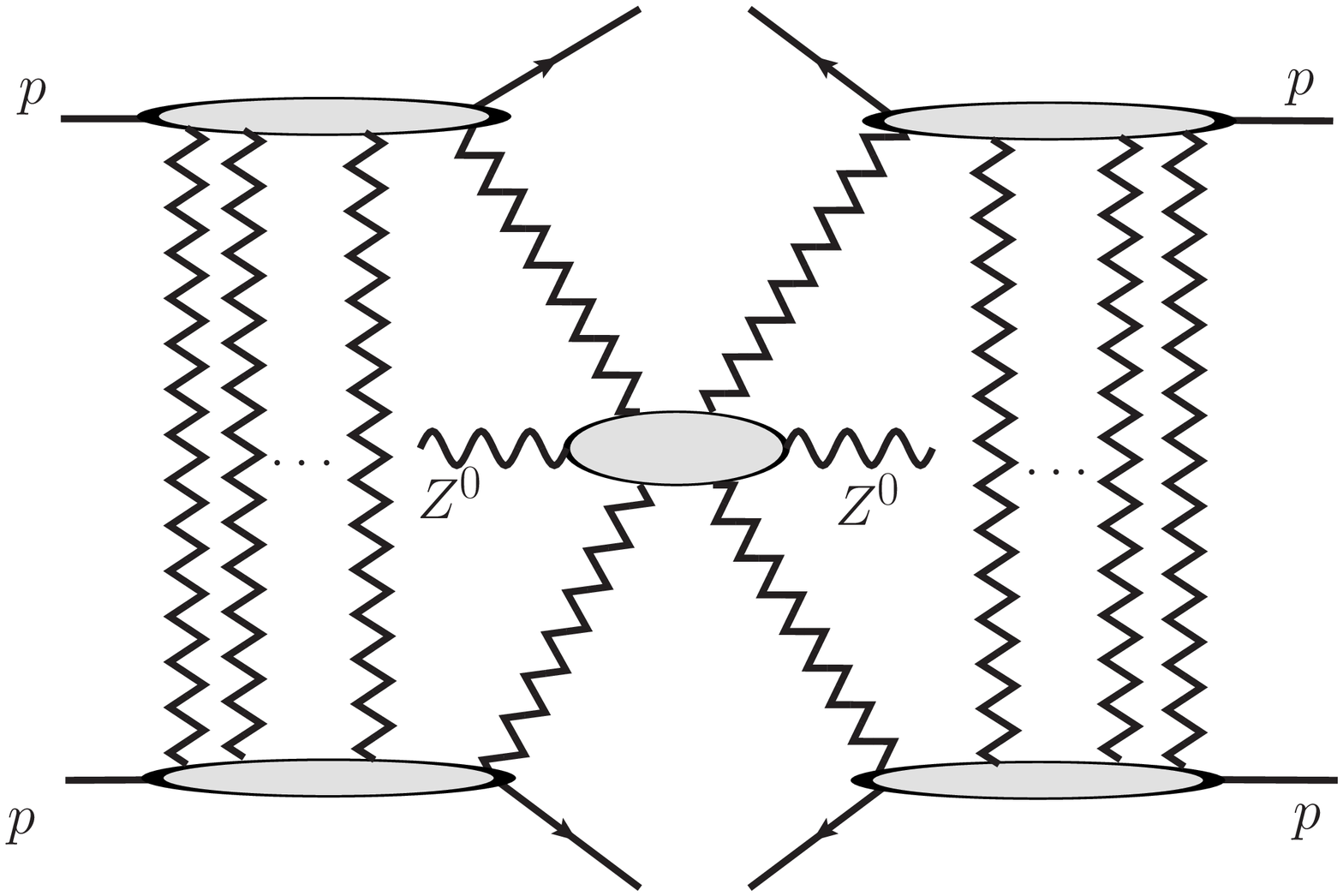}
\caption[*]{Absorptive correction to the $\Pom$--exchange 
contribution.
\label{fig:diagrams_absorption}
}
\end{center}
\end{figure}


It is however obvious, that the naive factorization prescription
of Eq.(\ref{naive_factorization}) cannot be correct. 
It neglects rescattering effects of incoming protons 
shown in Fig.\ref{fig:diagrams_absorption}. 
Indeed such diagrams quantify the probability that protons 
emerge intact out of the interactions region
in a regime where the typical events are highly inelastic 
and many channels are open \cite{Bjorken}. 
Here we restrict ourselves to only a 'minimal' 
scenario of factorization breaking induced by eikonalized 
multiple scatterings, following the formalism of 
Ref.\cite{TerMartirosyan} (for a somewhat modernized version, 
see \cite{KMR_eikonal}).

We do not enter here the debate on the possible relevance of 
multiple--Pomeron vertices (see for example the reviews 
\cite{Martin_epiphany,Maor} and references therein), 
but keep in mind that our treatment
of absorption may require a revision after better knowledge 
on soft interactions at the LHC has been acquired.

The relevant formulas are most easily written in 
impact parameter space. As in practice 
$t_i \sim -\bq_i^2$, where $\bq_i$ is the transverse momentum 
transfer to proton $i$, we can write
\begin{eqnarray}
f_\Pom (x_\Pom,t) = f_\Pom(x_\Pom, \bq) = f_\Pom(x_\Pom) \exp[-B \bq^2] \, ,
\end{eqnarray}
where $B=B(x_\Pom) = B_D + 2 \alpha'_\Pom \log(1/x_\Pom)$ is 
the $x_\Pom$--dependent diffractive slope. We follow the $H1$-analysis
\cite{H1}, and use the central values of their 
fit $B_D = 5.5 \, \mathrm{GeV}^{-2}
, \alpha'_\Pom =0.06 \, \mathrm{GeV}^{-2}$
Then, in impact parameter space, we have
\begin{eqnarray}
f_\Pom(x_\Pom,\bb) = \int {d^2 \bq \over (2\pi)^2} \, f_\Pom(x_\Pom,\bq) \, 
\exp[-i \bb \bq] = {f_\Pom(x_\Pom) \over 4 \pi B} \exp[- {\bb^2 \over 4 B} ] 
\equiv f_\Pom(x_\Pom) \, t_\Pom(x_\Pom,\bb) \, .
\nonumber \\ 
\label{impact_parameter_flux}
\end{eqnarray}
Here
\begin{eqnarray}
f_\Pom(x_\Pom) = f_\Pom(x_\Pom,\bq = 0) = \int d^2 \bb f_\Pom(x_\Pom,\bb) \, .
\end{eqnarray}
Now, we should make the following replacement in the cross section:

\begin{eqnarray}
f_\Pom(x_{\Pom,1}) \, f_\Pom(x_{\Pom,2}) &&= 
\int d^2\bb d^2 \bb_1 d^2 \bb_2 \delta^2(\bb - \bb_1 + \bb_2) \,  
f_\Pom(x_{\Pom,1},\bb_1) f_\Pom(x_{\Pom,2}, \bb_2) 
\nonumber \\
&&\longrightarrow \int d^2\bb d^2 \bb_1 d^2 \bb_2 \delta^2(\bb -\bb_1 + \bb_2) \, S^2_{abs}(\bb)  
f_\Pom(x_{\Pom,1},\bb_1) f_\Pom(x_{\Pom,2}, \bb_2) \, .
\nonumber
\end{eqnarray}

The treatment of absorptive corrections is in fact fully analogous
to the one required for $\gamma \gamma$ collisions in heavy ion collisions,
compare e.g. Eq.(2.2) in Ref. \cite{KSS}.

Using the form (\ref{impact_parameter_flux}) of the Pomeron--flux,
we observe, that the Born--level cross section will be multiplied
by the effective survival probability factor
\begin{eqnarray}
\overline{S^2}(x_{\Pom,1},x_{\Pom,2}) &&= 
\int d^2\bb d^2 \bb_1 d^2 \bb_2 S^2_{abs}(\bb) \delta^2 (\bb-\bb_1 + \bb_2)
t_\Pom(x_{\Pom,1}, \bb_1) t_\Pom(x_{\Pom,2},\bb_2) \, \nonumber \\
&&= {1 \over 4 \pi B_{12}(x_{\Pom,1},x_{\Pom,2})} 
\int d^2 \bb S^2_{abs}(\bb) 
\exp\Big[- {\bb^2 \over 4 B_{12}(x_{\Pom,1},x_{\Pom,2})} \Big] 
\, .
\label{gap_survival}
\end{eqnarray}
Here $B_{12}(x_{\Pom,1},x_{\Pom,2}) = 
B(x_{\Pom,1})B(x_{\Pom,2})/(B(x_{\Pom,1})+B(x_{\Pom,2}))$.
In fact, due to the very small Pomeron Regge-slope in the H1--fit
the $x_{\Pom,1}-x_{\Pom,2}$ dependence can be safely neglected.
A two--channel model for the absorption factor $S_{abs}^2(\bb)$,
is described in the appendix. 
It yields the numbers given in Table \ref{table:survival_probability}.

\begin{table}
\caption{Survival probability factor for purely exclusive 
and inclusive double diffractive processes.}
\label{table:survival_probability}
\begin{ruledtabular}
\begin{tabular}{|c|c|c|}
\hline
$\sqrt{s} [\mathrm{GeV}]$ & $pp \to p + Z^0 X +p$ & $pp \to Y_1 + Z_0 X + Y_2$ \\
\hline
1960  & $\sim 0.05$   & $\sim 0.06$ \\
\hline
14 000 & $\sim 0.025$ & $\sim 0.03$ \\
\hline
\end{tabular}
\end{ruledtabular}
\end{table}

\section{Results}

\begin{figure}[!htb] %
\begin{center}
\includegraphics[height=8cm]{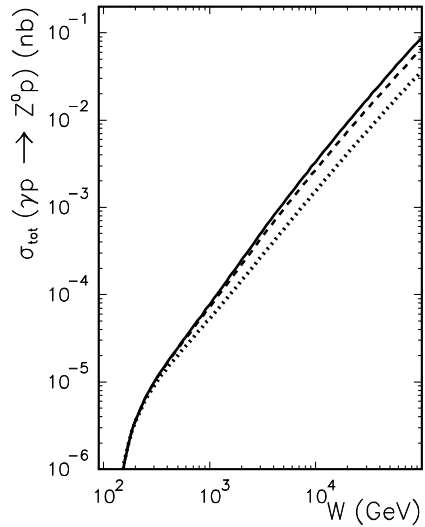}
\caption[*]{
The total cross section for $\gamma p \to Z^0 p$ reaction
as a function of photon-proton center-of-mass energy
for the Ivanov-Nikolaev UGDF. The dotted line includes:
u+d+s, the dashed line u+d+s+c and the solid line:
u+d+s+c+b.
\label{fig:gammap_Z0p}
}
\end{center}
\end{figure}

Before we go to hadronic reactions let us first present
predictions for the $\gamma p \to Z^0 p$ reaction.
In Fig.\ref{fig:gammap_Z0p} we show the total 
cross section as a function of photon-proton
center of mass energy $W$.
In this calculation we have used the 
unintegrated gluon distribution from Ref. \cite{IN_glue} 
and the slope parameter $B$ taken from Eq.(\ref{slope_exclusive}).
The cross section grows quickly with the energy
from the kinematical threshold $W_{th} = m_Z + m_p$.
At typical HERA energy $W$ = 200 GeV the cross section
is of the order of 10$^{-5}$ nb, i.e. too small to 
be measured. However, it grows quickly with energy
and at $W$ = 10 TeV it is already of the order of 1 pb.
We show not only the cross section with the full
amplitude (including all flavours) but also results
with three (u+d+s: dotted), four (u+d+s+c: dashed)
and five (u+d+s+c+b: solid) flavours. At low energies
it is enough to include only light flavours, while
at large energies all flavours must be included.


\begin{figure}[!htb] %
\begin{center}
\subfigure[]{\label{a}
\includegraphics[height=5cm]{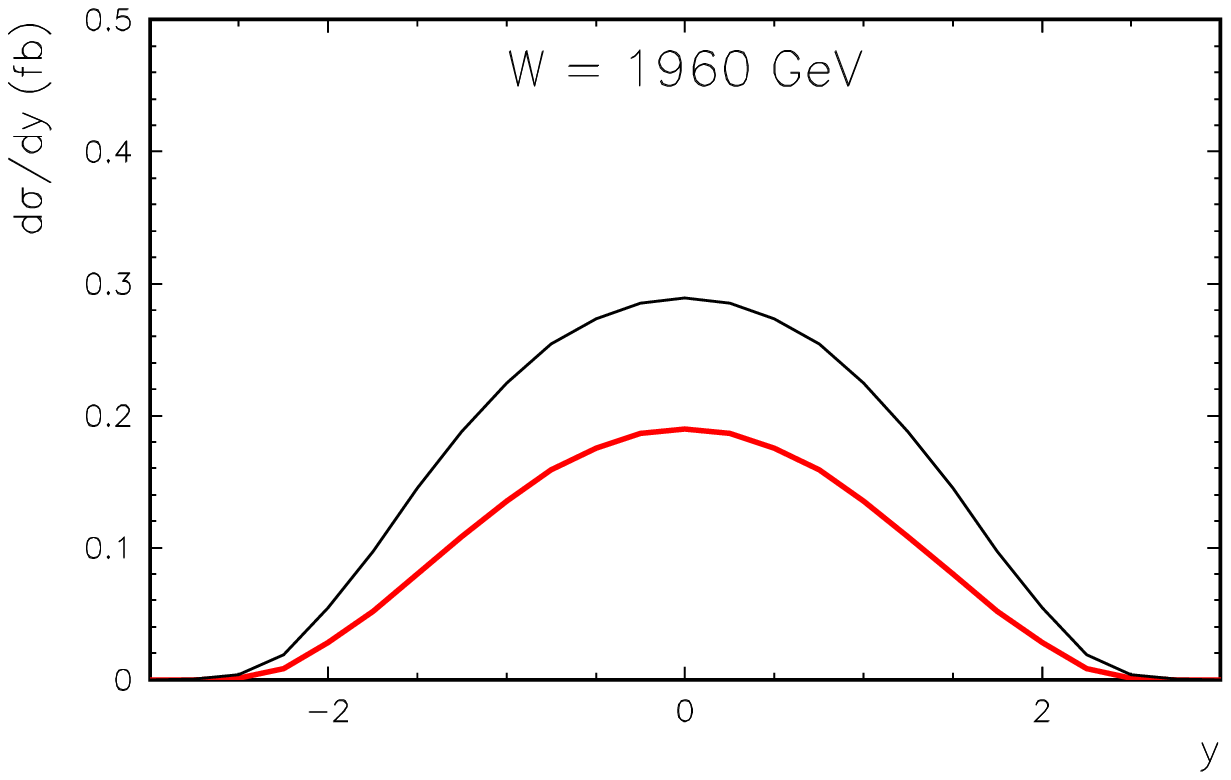}}
\subfigure[]{\label{b}
\includegraphics[height=5cm]{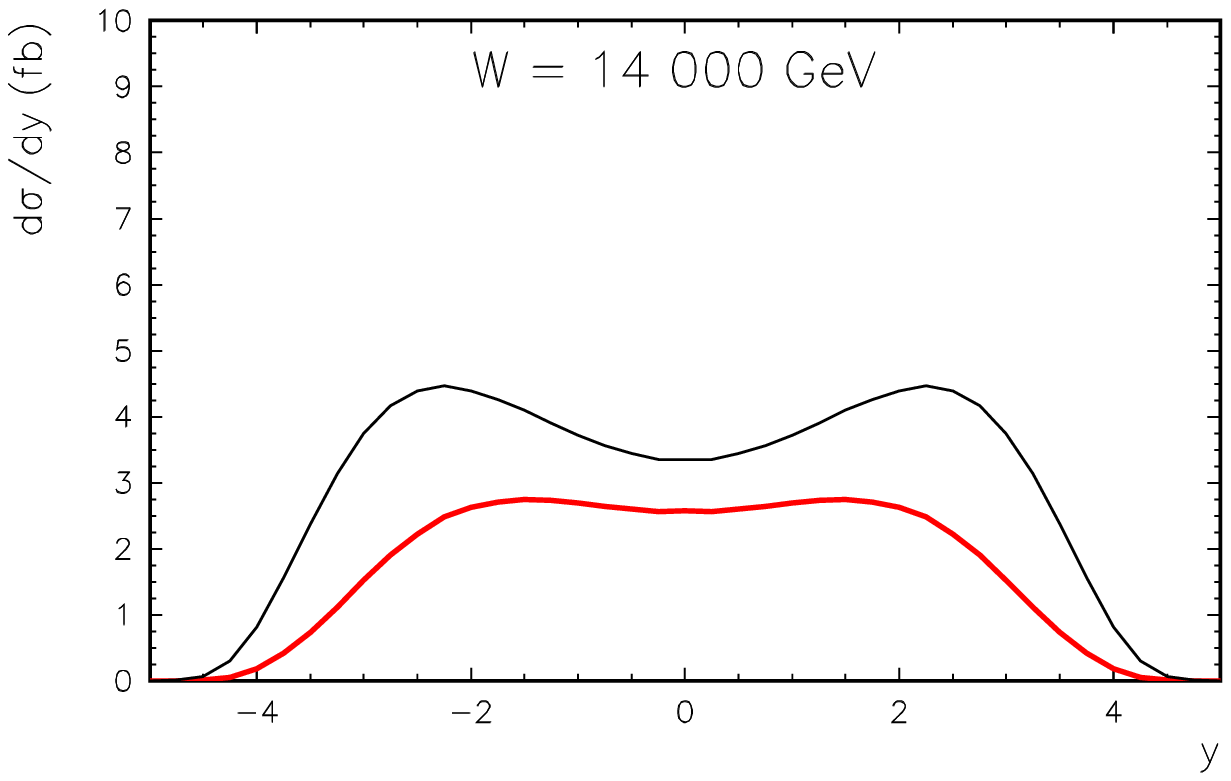}}
\caption[*]{Rapidity distribution of the exclusively 
produced $Z^0$ for the Tevatron (left) and LHC (right) energies.
The black thin solid line corresponds to Born amplitudes and 
the grey thick solid line (red on-line) includes absorption 
effects. 
\label{fig:dsig_dy}
}
\end{center}
\end{figure}

\begin{figure}[!htb] %
\begin{center}
\subfigure[]{\label{a}
\includegraphics[height=5cm]{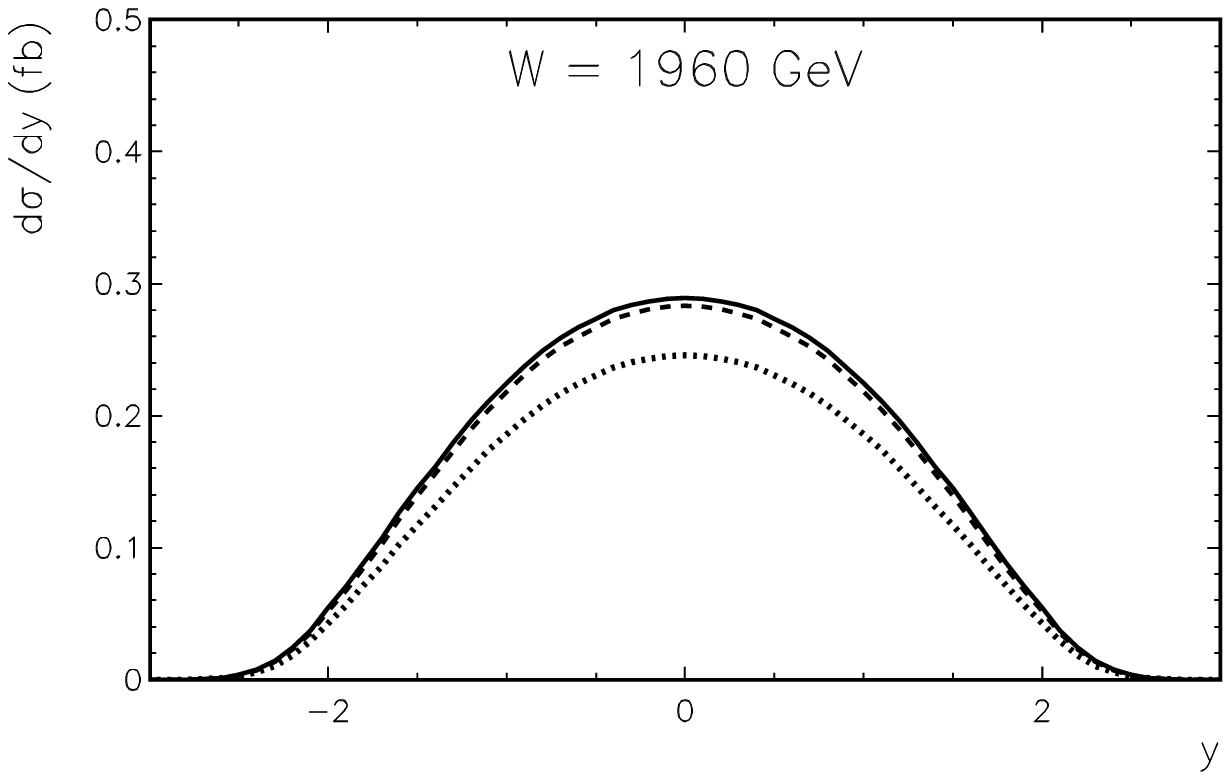}}
\subfigure[]{\label{b}
\includegraphics[height=5cm]{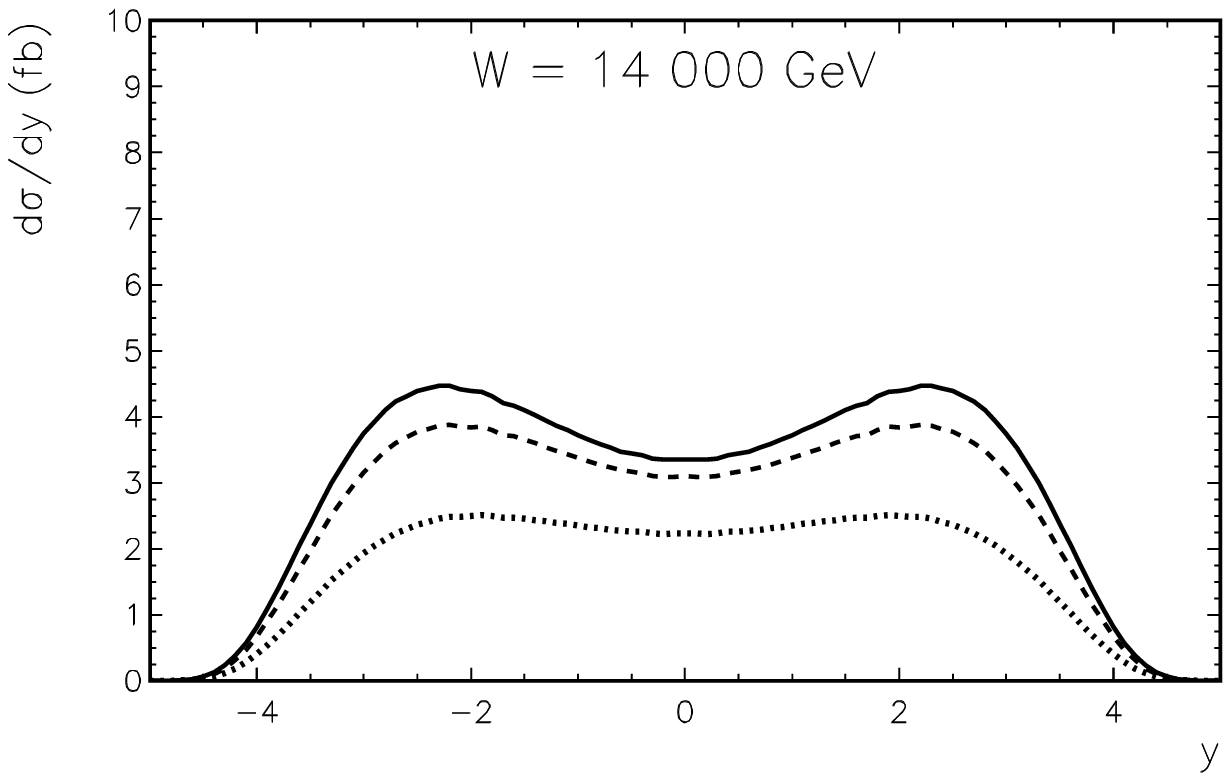}}
\caption[*]{
The influence of different flavours on rapidity
distributions of $Z^0$ for Tevatron (left) and LHC (right) 
energies. The dotted line includes:
u+d+s, the dashed line u+d+s+c and the solid line:
u+d+s+c+b.
\label{fig:dsig_dy_flavour_components}
}
\end{center}
\end{figure}

In Fig.\ref{fig:dsig_dy} we show distributions
in rapidity for the $p \bar p \to p \bar p Z^0$ (Tevatron)
and $p p \to p p Z^0$ (LHC) without (black thin solid) and
with (grey thick solid) absorption effects.
The Born approximation cross section calculated here
is much larger than that calculated in the dipole approach
in Refs.\cite{GM08,MW08}. 
Generally the absorption effects lower the cross section.
The effect depends on the rapidity.
Absorptive corrections for exclusive $Z^0$ production 
are bigger than for the exclusive production of 
$J/\Psi$ \cite{SS07} and 
$\Upsilon$ \cite{RSS08}. This is due to the fact that
for heavy particle production on average higher
four momentum transfers (and hence less peripheral collisions)
are involved than for lighter particles. 
Analogously as for photoproduction $\gamma p \to Z^0 p$
in Fig.\ref{fig:dsig_dy_flavour_components} we show 
the distribution in $Z$-boson rapidity in 
the Born approximation calculated
with different number of flavours included in the calculation.
While for the Tevatron energy it is enough
to include four flavours (u,d,s,c) at the LHC energy 
five flavours must be included. At LHC the inclusion of the 
$b$ quarks increases the cross section by about 20 \%.

\begin{figure}[!h]    %
\includegraphics[width=0.45\textwidth]{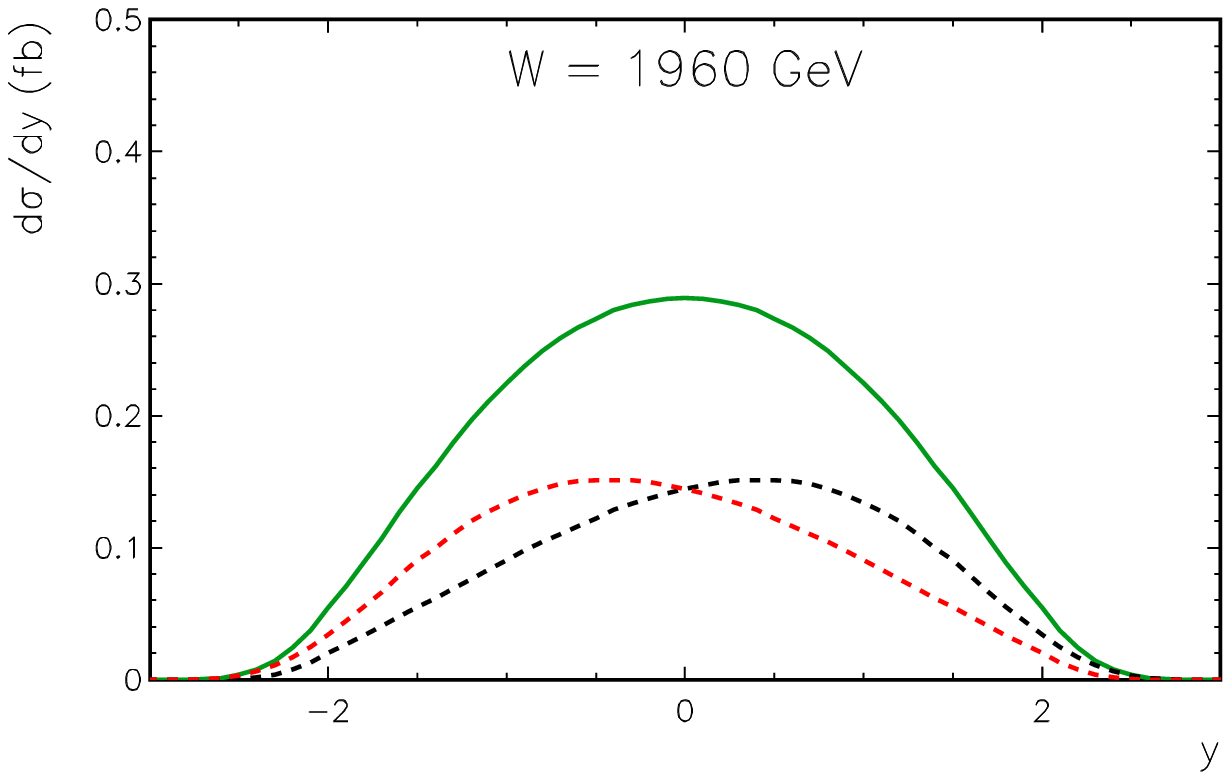}
\includegraphics[width=0.45\textwidth]{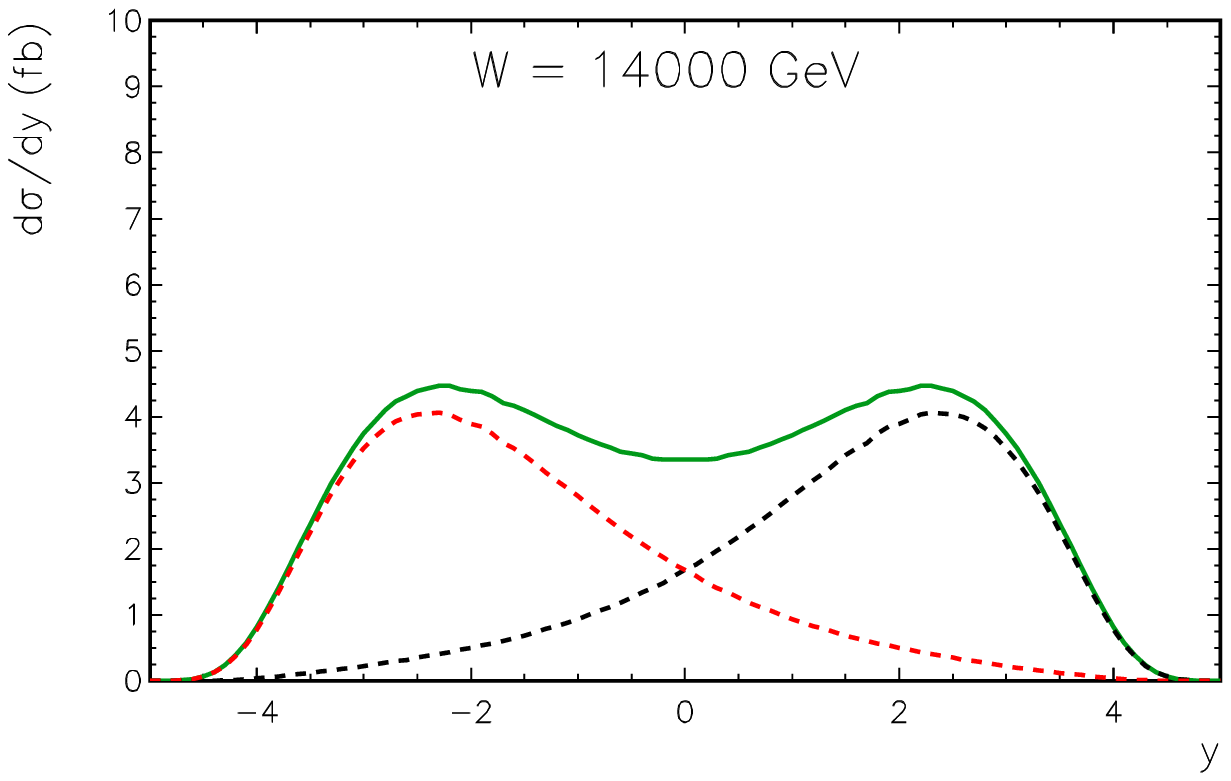}
   \caption{\label{fig:dsig_dy_components}
   \small The photon-pomeron and pomeron-photon
contributions for the Tevatron (left) and LHC (right) energies. 
No absorptive corrections are included here.}
\end{figure}

In Fig.\ref{fig:dsig_dy_components} we show separate
contributions of photon-pomeron and 
pomeron-photon fusion mechanisms as well as 
the sum of both processes. We wish to stress the fact
that in rapidity distributions all interference phenomena
dissapear if absorption corrections are neglected.
At LHC the two contributions are better separated
which leads to the camel-like shape
with minimum of the cross section at $y \approx$ 0.


The cross sections at the Tevatron energy 
$\sqrt{s}$ = 1.96 TeV is rather small. Recent search
for the exclusive $Z^0$ production \cite{CDF_Z0_exclusive} 
has found only upper limit for this process. 
There is a hope that at the LHC it could be measurable. 
One should remember, however, that in practice one 
can measure either $\mu^+ \mu^-$ or $e^+ e^-$ pairs. 
This means one can expect a sizeable background
from the $\gamma^* \gamma^* \to l^+ l^-$ processes
\cite{CDF_Z0_exclusive}
\footnote{Within the Standard Model, 
the $\gamma \gamma \to Z^0$ transition is absent
at the tree level. In fact, single-$Z^0$ production
in $\gamma \gamma$ collisions 
has up to now not been observed experimentally.}
. 
In order to get rid of this type of the background 
some cuts could be helpful.

\begin{figure}[!h]    %
\includegraphics[width=0.45\textwidth]{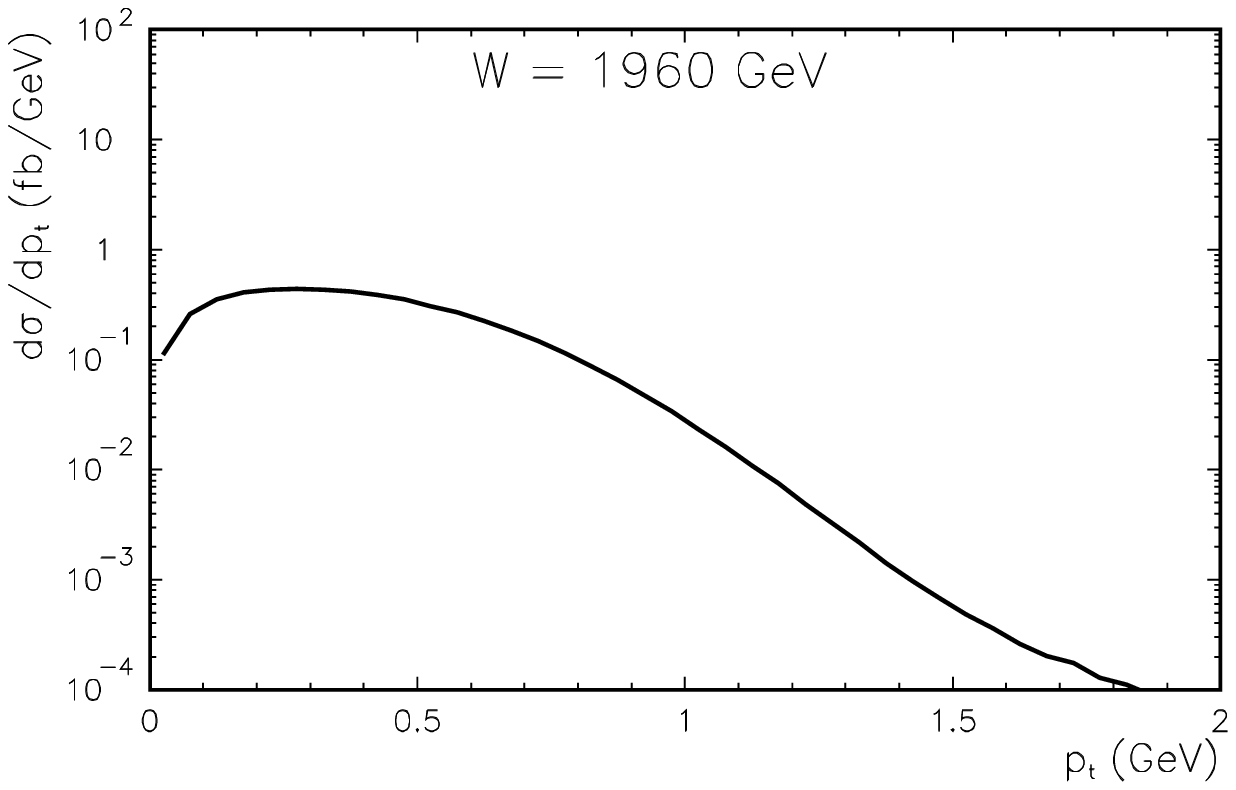}
\includegraphics[width=0.45\textwidth]{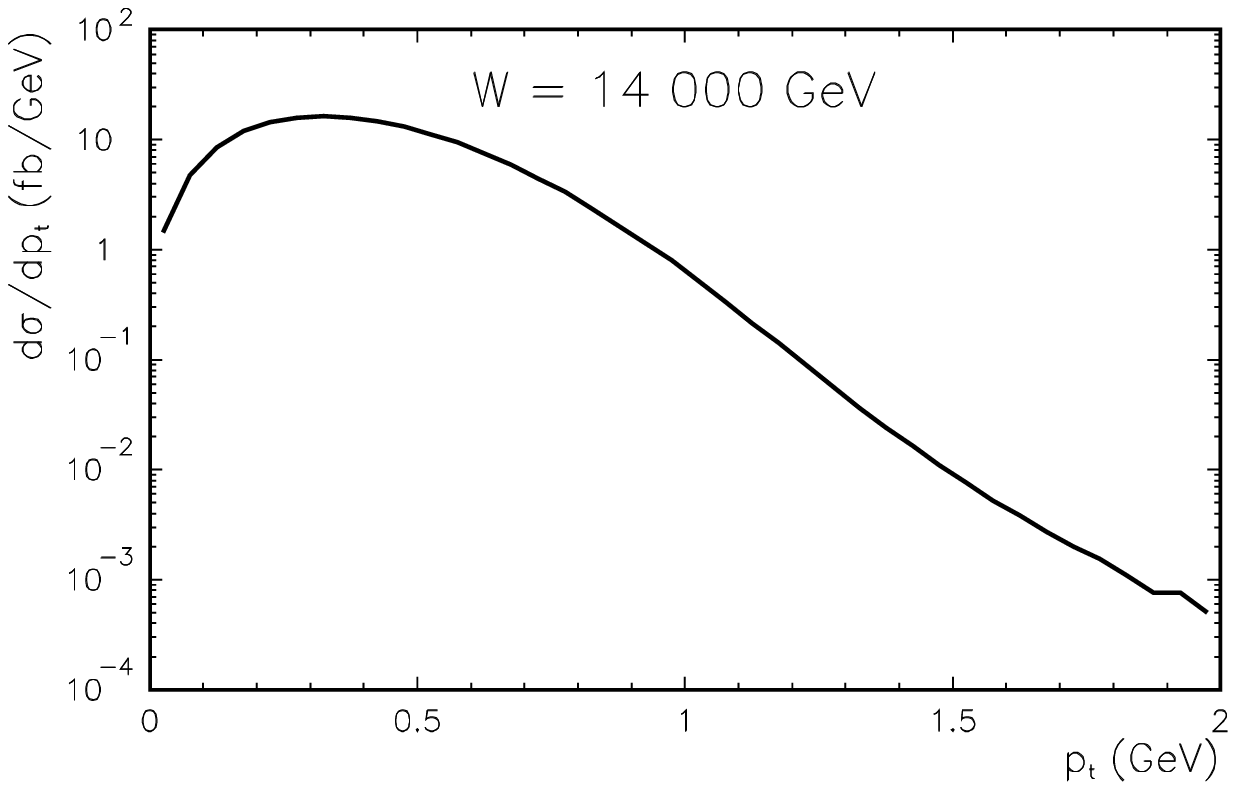}
   \caption{\label{fig:dsig_dpt}
   \small The transverse momentum distribution
of $Z^0$ bosons at $\sqrt{s}$ = 1960 GeV (left) and 
$\sqrt{s}$ = 14 TeV (right).
}
\end{figure}

In Fig.\ref{fig:dsig_dpt} we show transverse 
momentum distribution of the exclusively produced $Z^0$.
The distribution peaks at $p_t \sim$ 0.3 GeV and extends
to relatively large transverse momenta. This is in clear
contrast to the photon-photon processes where 
the corresponding transverse momenta of the lepton pair 
would peak at much lower transverse momenta. 
Imposing a lower 
cut on the lepton pair transverse momenta would cut off 
the unwanted photon-photon background.


\begin{figure}[!h]    %
\includegraphics[width=0.45\textwidth]{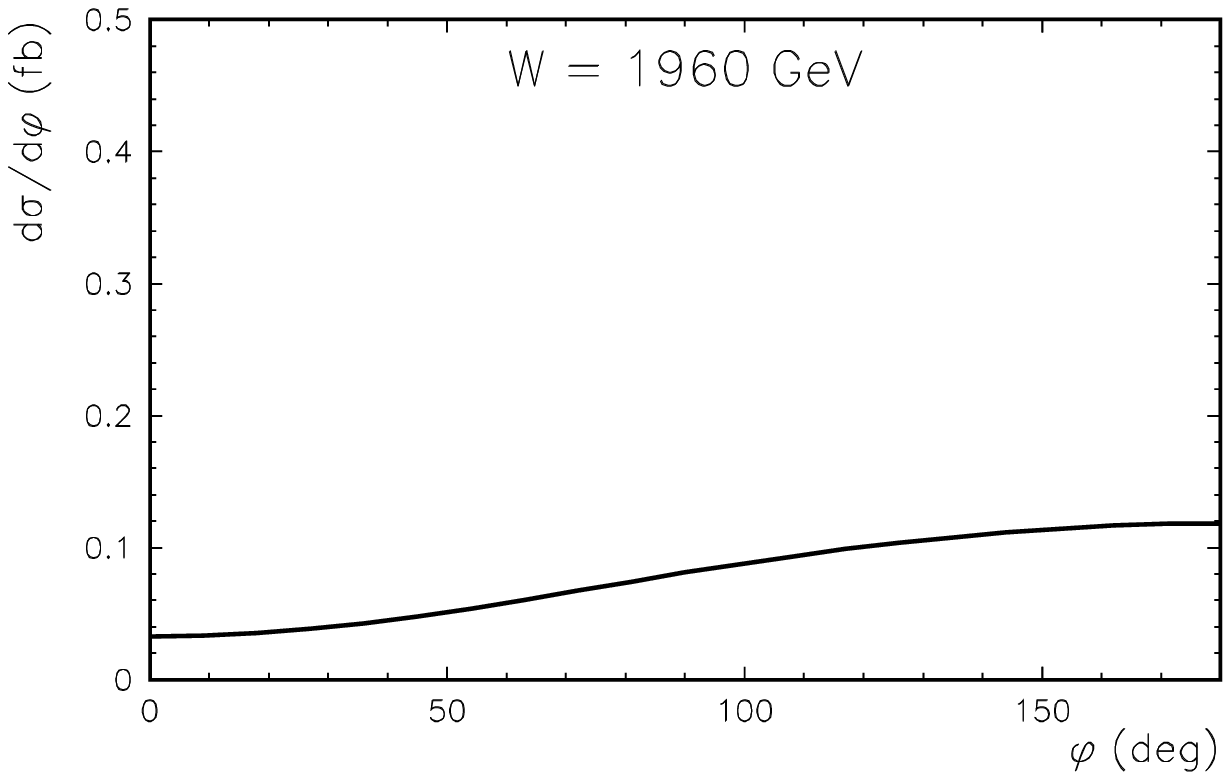}
\includegraphics[width=0.45\textwidth]{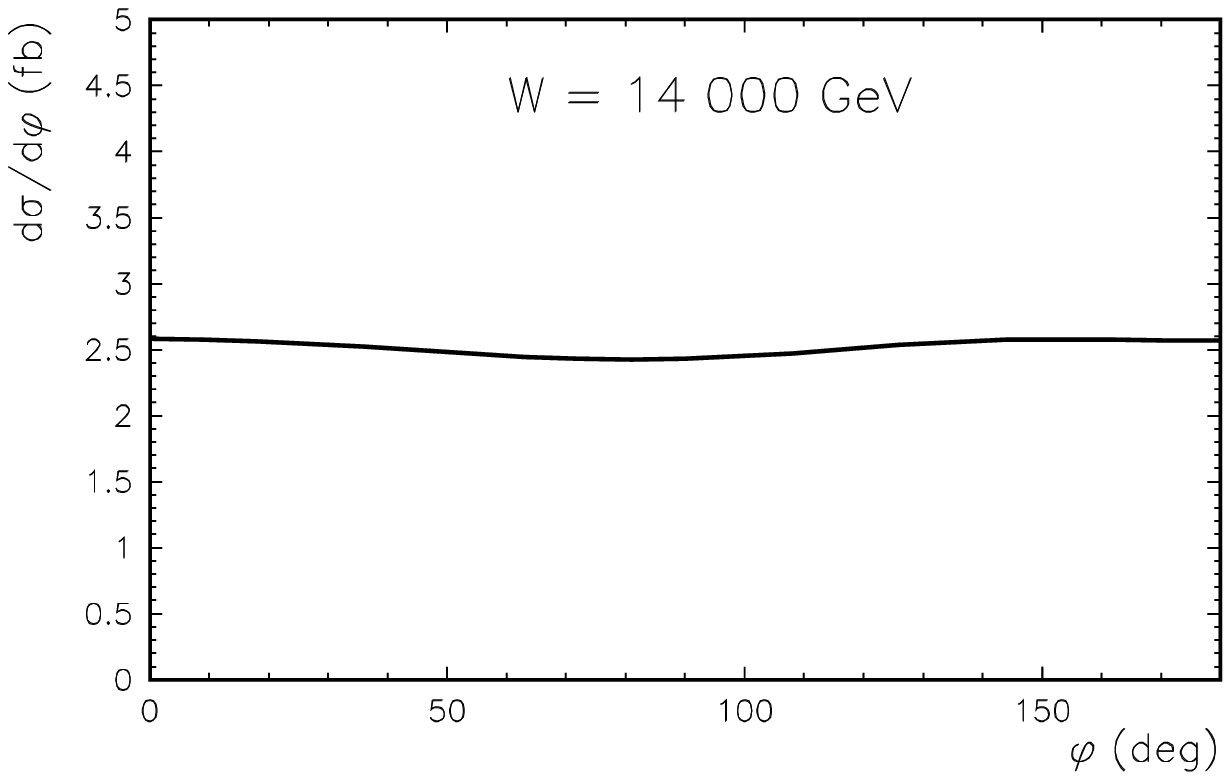}
   \caption{\label{fig:dsig_dphi}
   \small Distribution in relative azimuthal angle 
between outgoing protons at $\sqrt{s}$ = 1960 GeV (left)
and $\sqrt{s}$ = 14 TeV (right).
}
\end{figure}

There are definite plans that both ATLAS and CMS 
main detectors will be supplemented by several 
forward detectors.
In principle, having two forward detectors could allow 
to measure two outgoing protons in coincidence.
This could allow studying correlations between
outgoing protons.
As an example in Fig.\ref{fig:dsig_dphi} we show 
the distribution in relative azimuthal angle between 
outgoing protons.
Quite different distributions are obtained for the Tevatron 
($\bar p p$ collisions) and LHC ($pp$ collisions).
This effect is of the interference nature
and was already discussed for exclusive $J/\Psi$ production
\cite{SS07}.
In contrast to the reaction with $Z^0$ boson 
the relative azimuthal angle
distribution for the photon-photon processes peaks
sharply at $\phi \sim$ 180$^o$ \cite{CDF_Z0_exclusive}. 
Therefore imposing extra cuts in the azimuthal angle 
space should further diminish the photon-photon 
background opening a chance
to measure for the first time the 
exclusive $Z^0$ production in proton-proton collisions.


\begin{figure}[!htb] %
\begin{center}
\includegraphics[height=7.5cm]{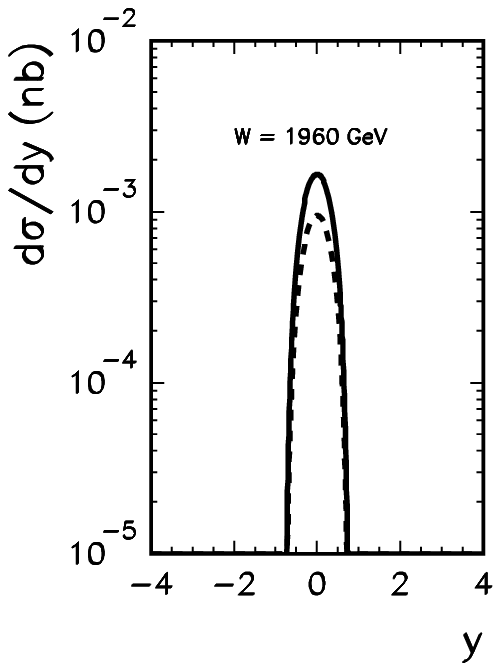}
\includegraphics[height=7.5cm]{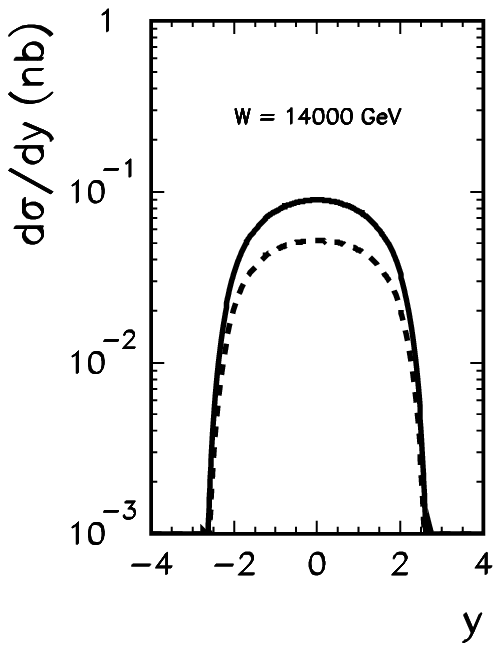}
\caption[*]{Rapidity distributions of inclusive 
double-pomeron production of $Z^0$ boson for the Tevatron
(left) and LHC (right). The solid lines include both pomeron
and reggeon exchanges while the dashed lines only pomeron exchanges.
In this calculation the fit (a) from Ref.\cite{H1} was used.
No absorptive corrections were included here.
\label{fig:dsig_dy_double_pomeron}
}
\end{center}
\end{figure}

Finally let us present our estimate of the inclusive
double-diffractive contribution of 
Fig.\ref{fig:double_pomeron_diagram}.
In Fig.\ref{fig:dsig_dy_double_pomeron} we show the
cross section with pomeron exchanges only (dashed) and with
both pomeron and reggeon exchanges included (solid).
This cross sections have to be multiplied in addition
by the gap survival probabilities from Table 1. 
In this calculation $x_{\Pom}^{max}$ = 0.1 was assumed.
This means cuts on longitudinal momentum fractions
of outgoing protons/antiprotons.
Even after including absorption corrections the inclusive
double-pomeron contribution is a few orders of magnitude
larger than the purely exclusive cross section.
The rapidity distributions from 
Fig.\ref{fig:dsig_dy_double_pomeron} are more narrow 
than the purely exclusive distributions shown earlier. 
This is partially related to the cuts on longitudinal 
momentum fractions and is of purely kinematic origin.
The cross section for inclusive double pomeron 
contribution is certainly measurable at LHC and is bigger
than the cross section for single-diffractive $Z^0$ production
at Tevatron \cite{D0}.


\begin{figure}[!htb] %
\begin{center}
\includegraphics[height=6.5cm]{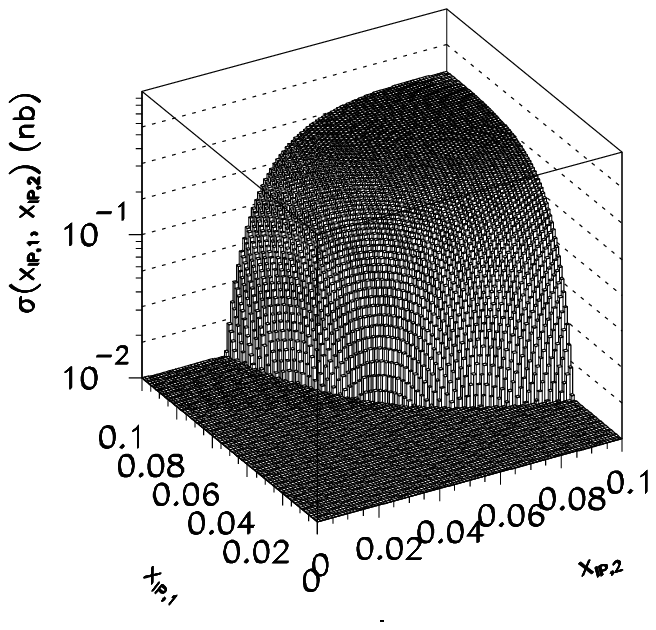}
\includegraphics[height=6.5cm]{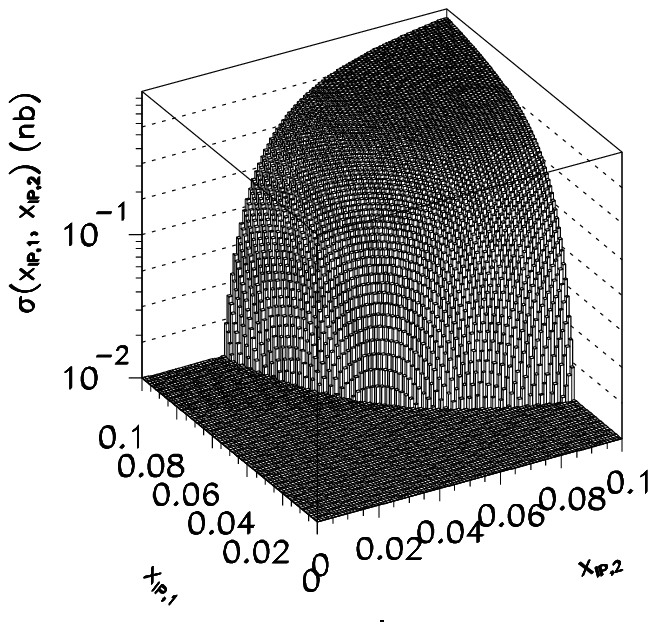}
\includegraphics[height=6.5cm]{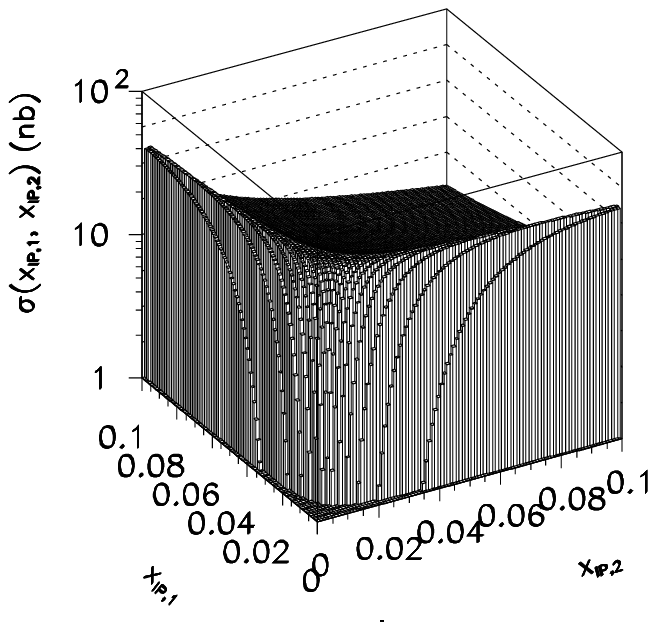}
\includegraphics[height=6.5cm]{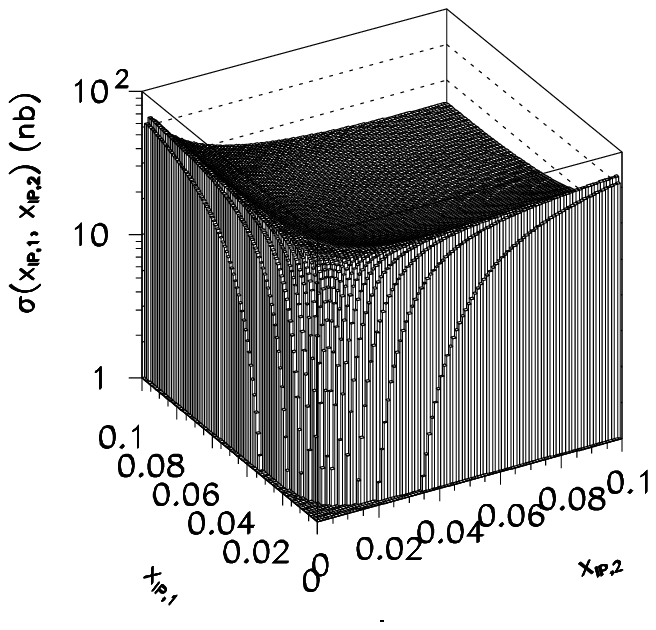}
\caption[*]{Two dimensional distributions in 
($x_{\Pom,1}, x_{\Pom,1}$) for the Tevatron energy W = 1960 GeV
(upper panels) and for the LHC energy W = 14000 GeV (lower panels).
The left panels include only pomeron exchanges while the right
panels both pomeron and reggeon exchanges.
In this calculation the fit (a) from Ref.\cite{H1} was used.
No absorbtion corrections were included here.
\label{fig:dsig_dx1dx2_double_pomeron}
}
\end{center}
\end{figure}

In Fig.\ref{fig:dsig_dx1dx2_double_pomeron} we show 
two-dimensional distribution in pomeron/reggeon momentum fractions 
($x_{\Pom,1}, x_{\Pom,2}$). The large mass of the $Z^0$ boson
causes that the small values of $x_{\Pom,1}$ and $x_{\Pom,2}$
are not accessible kinematically. This is more evident for
the Tevatron energy. This is also the reason for much smaller
cross section for double-diffractive production of $Z^0$ for
Tevatron compared to LHC.


\section{Conclusions}

We have extended the $k_\perp$-factorization to the 
exclusive production of $Z^0$ bosons.
The production amplitude was calculated
using an unintegrated gluon distribution \cite{IN_glue}
adjusted to inclusive deep inelastic structure functions.
The so obtained $\gamma p \to Z p$ amplitude served as 
an input for the evaluation of the $pp \to pp Z^0$ process. 
Our results obtained with bare
(i.e. without absorption) amplitudes are by a factor of 3 larger 
than those obtained earlier in the dipole approach. 
We have analyzed the role of individual flavours.
For low energy it is enough to include only light
flavours while at high energies all flavours must
be included.

Compared to earlier works in the literature we have 
taken into account absorption effects.
The absorption effects depend on the Z-boson rapidity and 
lower the cross section by a factor of 1.5-2.
As for the exclusive $\Upsilon$ production \cite{RSS08} 
the larger rapidity, the larger the absorption effect.

Very small cross sections are obtained both for
Tevatron and LHC. This means that possible background
must be studied.
The $Z^0$ is measured via $e^+ e^-$ or $\mu^+ \mu^-$
decay channels.
Recently the CDF collaboration \cite{CDF_Z0_exclusive}
presented a first estimate of the upper limit
for the exclusive $Z^0$ production at the Tevatron.
Their limit is about three orders of magnitude larger than
our predictions. This demonstrates difficulties
to measure the exclusive process. The situation will
improve at LHC (larger cross section, larger luminosity), 
but even there it will be rather difficult to measure 
the cross section. 

In a detailed analysis a background from 
the $\gamma \gamma \to l^+ l^-$
and $\gamma \Pom \to \gamma^* \to l^+ l^-$ (sub)processes
must be taken into account.  
We have found that distributions in $Z^0$ (lepton pair)
transverse momentum as well as in relative azimuthal angle
between outgoing protons can be very useful to separate out
the background $\gamma \gamma \to l^+ l^-$ processes.

To our knowledge for the first time in the literature,
we have estimated inclusive double diffractive production 
of $Z^0$ using diffractive parton distributions obtained recently
from the analysis of the proton diffractive structure
functions/diffractive dijets performed
by the H1 collaboration at HERA. We have calculated
cross section assuming Regge factorization as well as
inculding absorption effects leading to the factorization
breaking. Rather large inclusive double-diffractive cross
sections were found at LHC. A future experiments with
forward instrumentation of main LHC detectors 
(ATLAS and ALICE) should provide new results 
concerning hard diffraction.
This will allow to further investigate the mechanism of
Regge-factorization breaking observed already
for soft total single and double diffraction. 

\vspace{1cm}
\section{Acknowledgements}
We are indebted to Christophe Royon and Laurent Schoeffel 
for providing us with the H1 parton distributions in the pomeron.
This work was partially supported by the Polish Ministry 
of Science and Higher Education (MNiSW) under contracts 
MNiSW N N202 249235 and 1916/B/H03/2008/34.

\section{Appendix: Two-channel model for the absorptive corrections}

Here we present the details of the two-channel model used 
for the evaluation of absorptive corrections to central inclusive 
$Z^0$--production. It improves over a single--channel
desription by taking into account some inelastic 
shadowing corrections.
As physical states, one would include the proton $\ket{p}$ 
and some effective
low--mass states $\ket{N_i^*}$ with proton--quantum numbers 
(representing e.g. resonances and the diffractively 
excited $N\pi,N\pi \pi,...$ components.
The multichannel eikonal will then be an operator acting on the 
tensor--product space of physical states 
$\ket{N_i N_j} = \ket{N_i} \otimes \ket{N_j}$, where
$N_i = p,N^*$.
The proton--proton $S$--matrix in this space is now given by
\begin{eqnarray}
\hat{S}(\bb) = \exp\Big[-{1 \over 2} \nu(s,\bb) \, \hat{g} \otimes \hat{g} \Big]
\, ,
\end{eqnarray}
where the opacity $\nu(s,\bb)$ is given by
\begin{eqnarray}
\nu(s,\bb) = g_{pp}^2 \, 
\Big({s\over s_0}\Big)^{\Delta_\Pom} \, T_\Pom(\bb) \, , 
\label{eq:opacity}
\end{eqnarray}
and we stick to an oversimplified model, in which all matrix elements
have the same $\bb$ dependence given by $T_\Pom(\bb)$.
The $\bb$-space profile was taken in the Gaussian form:
\begin{eqnarray}
T_\Pom(\bb) = {1 \over 2 \pi (B_0 + 2 \alpha' \log(s/s_0))} 
\exp \Big[ {-\bb^2 \over 2 (B_0 + 2 \alpha' \log(s/s_0) )} \Big]
\; .
\end{eqnarray}
The values of the bare pomeron parameters 
$g_{pp}^2, \Delta_\Pom$ as well as the parametrisation of $T_\Pom(\bb)$
are given below.
Notice that below, we do not distinguish between protons 
and antiprotons, and will always refer to proton--proton
scattering even when discussing results for the Tevatron.

In the two--channel case, where all inelastic excitations
are subsumed in a single effective state $\ket{N^*}$, 
the matrix $\hat{g}$ is written as
\begin{eqnarray}
\hat{g} = 
\left(
\begin{array}{cc}
1 + \delta & \gamma \\
\gamma & 1 - \delta 
\end{array}
\right) \, . 
\end{eqnarray}
It has the eigenvalues 
\begin{eqnarray}
\lambda_{1,2} =  1 \pm \sqrt{\delta^2 + \gamma^2} \, ,
\end{eqnarray}
and the physical states can be expanded into $S$--matrix eigenstates as
\begin{eqnarray}
\ket{p} = \sum_i C_i^p \ket{i} \, , \, \ket{N^*} = \sum_i C_i^{N^*} \ket{i} \, .
\label{eq:eigenstates}
\end{eqnarray}

Now we turn to the gap survival probabilty and evaluate the
effective $S^2_{abs}(\bb)$ which enters Eq.(\ref{gap_survival}).
We distinguish different final states:

\subsection{$p p \to p + Z^0 X + p$}
First let us the case of the proton--proton final state.
Here we need to substitute

\begin{eqnarray}
S^2_{abs}(\bb) \longrightarrow 
\Big| \bra{pp} \Big(\hat{g} \otimes \hat{g}\Big) 
\hat{S}(\bb) \ket{pp} \Big|^2 \, 
= \Big| \bra{pp} \Big(\hat{g} \otimes \hat{g}\Big) 
\exp[-{1 \over 2} \nu(s,\bb) \hat{g} \otimes \hat{g} ] \ket{pp} 
\Big|^2
\, .
\end{eqnarray}
To evaluate the matrix element $\bra{pp} \dots \ket{pp}$, we
should expand the protons into eigenstates of the $S$--matrix 
according to Eq.(\ref{eq:eigenstates}):
\begin{eqnarray}
\bra{pp} \Big(\hat{g} \otimes \hat{g}\Big) 
\hat{S}(\bb) \ket{pp} 
&&= |C^p_1|^4 \lambda_1^2 e^{-\nu \lambda_1^2/2}
+ |C^p_2|^4 \lambda_2^2 e^{-\nu \lambda_2^2/2} 
+ 2 \,  |C^p_1|^2 |C^p_2|^2  \lambda_1 \lambda_2 
e^{-\nu \lambda_1 \lambda_2/2} \, .
\nonumber \\
\end{eqnarray}
Here we suppressed the arguments of $\nu = \nu(s,\bb)$.
\subsection{$pp \to Y_1 + Z^0 X + Y_2$}
If protons in the final state cannot be measured, we need to 
sum over all excitations $Y_{1,2} \in \{p,N^*\}$, and we should 
substitute 
\begin{eqnarray}
S_{abs}^2(\bb) &\longrightarrow& \sum_{Y_1,Y_2} \Big| \bra{Y_1 Y_2} 
\Big( \hat{g} \otimes \hat{g} \Big)
\hat{S}(\bb) \ket{pp} \Big|^2 = \bra{pp} 
\Big( \hat{g} \otimes \hat{g} \Big)^2 \hat{S}^2(\bb) \ket{pp} 
\nonumber \\
& = &
\bra{pp} 
\Big( \hat{g} \otimes \hat{g} \Big)^2 
\exp[-\nu(s,\bb) \hat{g} \otimes \hat{g}] \ket{pp}
\nonumber \\
& = &
  |C^p_1|^4 \lambda_1^4 e^{-\nu \lambda_1^2} 
+ |C^p_2|^4 \lambda_2^4 e^{-\nu \lambda_2^2} 
+ 2 \,  |C^p_1|^2 |C^p_2|^2  (\lambda_1 \lambda_2)^2 
e^{-\nu \lambda_1 \lambda_2}
\, .
\end{eqnarray}
%
%
%
Equivalent equations can be found in \cite{KMR_eikonal},
who we largely follow in choosing $\gamma = 0.55, \delta = 0$.
Then
\begin{eqnarray}
\ket{p} = {1 \over \sqrt{2}}\Big( \ket{1} + \ket{2} \Big) 
\, , \, \ket{N^*} = {1 \over \sqrt{2}}\Big( \ket{1} - \ket{2} \Big) 
\, ,
\end{eqnarray}
with eigenvalues $ \lambda_{1,2} =   (1 \pm \gamma)$.
The bare Pomeron parameters used in the parametrisation of the opacity
(\ref{eq:opacity}) 
with $s_0 = 1 \, \mathrm{GeV}^2$,  are taken as
\begin{eqnarray}
g_{pp}^2 = 27 \, \mathrm{mb} \, , \, \Delta_\Pom = 0.11 \, , 
\, B_0 = 9 \, \mathrm{GeV}^{-2} \, ,\, \alpha' = 0.14 \, \mathrm{GeV}^{-2} \, 
\, .
\end{eqnarray}
These parameters are so adjusted, that we obtain reasonable values for
the total cross section $\sigma_{tot}$, the elastic cross section 
$\sigma_{el}$, as well as the elastic slope $B_{el}$.
They are obtained from
\begin{eqnarray}
\sigma_{tot} = 2 \,  \int d^2 \bb \, \Gamma(\bb) \,\, , \, 
\sigma_{el} = \int d^2 \bb \, \Gamma^2(\bb) \,\, , \, 
B_{el} = {1 \over 2} {\int d^2 \bb \, \bb^2 \Gamma(\bb) \over 
\int  d^2\bb \, \Gamma(\bb) } \, ,
\end{eqnarray}
where the impact--parameter space forward amplitude $\Gamma(\bb)$ is
given by
\begin{eqnarray}
\Gamma(\bb) = 1 - \bra{pp} \hat{S}(\bb) \ket{pp} \, .
\end{eqnarray}
For the energy of Tevatron Run I, $\sqrt{s} = 1800 \, \mathrm{GeV}$, we obtain
$\sigma_{tot} = 78.5 \, \mathrm{mb}$, $\sigma_{el} = 16.7 \, \mathrm{mb}$,
and $B_{el} = 17.2  \, \mathrm{GeV}^{-2}$. For the LHC energy of 
$\sqrt{s} = 14 \, \mathrm{TeV}$, this oversimplified model predicts
$\sigma_{tot} = 106 \, \mathrm{mb}$, $\sigma_{el} = 26 \, \mathrm{mb}$,
and $B_{el} = 19.8  \, \mathrm{GeV}^{-2}$. 




\begin{thebibliography}{100}


\bibitem{exclusive_dijets}
  T.~Aaltonen {\it et al.}  [CDF Collaboration],
  Phys.\ Rev.\  D {\bf 77}, 052004 (2008)
  [arXiv:0712.0604 [hep-ex]].

\bibitem{exclusive_charmonia}
  T.~Aaltonen {\it et al.}  [CDF Collaboration],
  arXiv:0902.1271 [hep-ex].

\bibitem{Martin_epiphany}
  A.~D.~Martin, M.~G.~Ryskin and V.~A.~Khoze,
  arXiv:0903.2980 [hep-ph].

\bibitem{experiment}
  M.~G.~Albrow {\it et al.}  [FP420 R and D Collaboration],
  arXiv:0806.0302 [hep-ex];
 R.~Schicker,
  AIP Conf.\ Proc.\  {\bf 1105}, 136 (2009)
  [arXiv:0812.3123 [hep-ex]].

\bibitem{SS07}
  W.~Sch\"afer and A.~Szczurek,
  Phys.\ Rev.\  D {\bf 76}, 094014 (2007).

\bibitem{RSS08}
  A.~Rybarska, W.~Sch\"afer and A.~Szczurek,
  Phys.\ Lett.\  B {\bf 668}, 126 (2008)
  [arXiv:0805.0717 [hep-ph]].

\bibitem{GM08}
  V.~P.~Goncalves and M.~V.~T.~Machado,
  Eur.\ Phys.\ J.\  C {\bf 56}, 33 (2008)
  [arXiv:0710.4287 [hep-ph]].

\bibitem{MW08}
  L.~Motyka and G.~Watt,
  Phys.\ Rev.\  D {\bf 78}, 014023 (2008)
  [arXiv:0805.2113 [hep-ph]].

\bibitem{INS06}
 I.~P.~Ivanov, N.~N.~Nikolaev and A.~A.~Savin,
  Phys.\ Part.\ Nucl.\  {\bf 37}, 1 (2006).


\bibitem{CDF_Z0_exclusive}
  T.~Aaltonen {\it et al.}  [CDF Collaboration],
  arXiv:0902.2816 [hep-ex].




\bibitem{Single_Diffraction}
P. Bruni and G. Ingelman, 
Phys. Lett. {\bf B311} (1993) 317;
L. Alvero, J.C. Collins, J. Terron and J.J. Whitmore,
Phys. Rev. {\bf D59} (1999) 074022.

\bibitem{Shuvaev}
  A.~G.~Shuvaev, K.~J.~Golec-Biernat, A.~D.~Martin and M.~G.~Ryskin,
  Phys.\ Rev.\  D {\bf 60}, 014015 (1999).

\bibitem{NZ91}
  N.~N.~Nikolaev and B.~G.~Zakharov,
  Z.\ Phys.\  C {\bf 49}, 607 (1991).

\bibitem{KNNZ93}
  B.~Z.~Kopeliovich, J.~Nemchick, N.~N.~Nikolaev and B.~G.~Zakharov,
  Phys.\ Lett.\  B {\bf 309}, 179 (1993)
  [arXiv:hep-ph/9305225].

\bibitem{IN_glue}
  I.~P.~Ivanov and N.~N.~Nikolaev,
  Phys.\ Rev.\  D {\bf 65}, 054004 (2002).


\bibitem{H1_JPsi}
A.~Aktas {\it et al.} [H1 Collaboration],
  Eur.\ Phys.\ J.\  C {\bf 46}, 585 (2006).

\bibitem{DDLN02}
S. Donnachie, G. Dosch, P. Landshoff and O. Nachtmann,
"Pomeron Physics and QCD", Cambridge University Press, Cambridge 2002.


\bibitem{CDF_total}
F.~Abe et al. [CDF Collaboration], Phys. Rev. {\bf D50}, 5518 (1994).

\bibitem{Ingelman_Schlein}
  G.~Ingelman and P.~E.~Schlein,
  Phys.\ Lett.\  B {\bf 152}, 256 (1985).

\bibitem{BP87}
V.~Barger and R.~Phillips, "Collider Physics", Addison-Wesley Publishing
Company, Redwood Cite, 1987.

\bibitem{H1}
  A.~Aktas {\it et al.}  [H1 Collaboration],
  Eur.\ Phys.\ J.\  C {\bf 48}, 715 (2006)
  [arXiv:hep-ex/0606004].

\bibitem{WS98}
  W.~Sch\"afer,
  arXiv:hep-ph/9806295, in: 
Deep Inelastic Scattering and QCD: DIS 98: Proceedings. Edited by Gh. Coremans and R. Roosen. Singapore, World Scientific, 1998.

\bibitem{Bjorken} 
  J.~D.~Bjorken,
  Phys.\ Rev.\  D {\bf 47}, 101 (1993).

\bibitem{TerMartirosyan}
  K.~A.~Ter-Martirosyan,
  Sov.\ J.\ Nucl.\ Phys.\  {\bf 10}, 600 (1970)
  [Yad.\ Fiz.\  {\bf 10}, 1047 (1969)].

\bibitem{KMR_eikonal}
  V.~A.~Khoze, A.~D.~Martin and M.~G.~Ryskin,
  Eur.\ Phys.\ J.\  C {\bf 18}, 167 (2000)
  [arXiv:hep-ph/0007359].

\bibitem{Maor}
  U.~Maor,
  AIP Conf.\ Proc.\  {\bf 1105}, 248 (2009)
  [arXiv:0811.2636 [hep-ph]].

\bibitem{KSS}
  M.~K\l usek, W.~Sch\"afer and A.~Szczurek,
  Phys.\ Lett.\  B {\bf 674}, 92 (2009)
  [arXiv:0902.1689 [hep-ph]].

\bibitem{D0}
  V.~M.~Abazov {\it et al.}  [D0 Collaboration],
  Phys.\ Lett.\  B {\bf 574}, 169 (2003)
  [arXiv:hep-ex/0308032].

\end{thebibliography}
\end{document}